# Heat transfer and mixing in initiated Chemical Vapor Deposition analyzed by in-situ gas composition sensing

Simon Shindler [1] and Rong Yang [2,a]

[1]Cornell University, School of Engineering, Robert Frederick Smith School of Chemical and Biomolecular Engineering, Olin Hall, Ho plaza, Ithaca, NY 14853
[2]Cornell University, School of Engineering, Robert Frederick Smith School of Chemical and Biomolecular Engineering, Olin Hall, Ho plaza, Ithaca, NY 14853

[a] Electronic mail: ryang@cornell.edu

Fundamental studies of iCVD kinetics often make assumptions about the iCVD vapor phase that have been challenging to validate experimentally. We address this gap by investigating heat transfer and mixing in the iCVD reactor using in-situ gas composition sensing. Our work allows practitioners of iCVD to estimate the degree of mixing and temperature profile in the vapor phase based on reactor dimensions and process variables such as chamber pressure. We begin by using dimensional analysis to simplify the parameter space governing heat transfer and mixing, identifying key dimensionless groups that capture the dominant physics. We find that the degree of mixing in the reactor is primarily determined by the Peclet number, and heat transfer is primarily determined by the Knudsen number (Kn). Multiphysics simulations of the iCVD reactor provide visualizations of non-ideal mixing and allow us to identify the critical Peclet number above which the well-mixed assumption may begin to break down. To verify that the well-mixed assumption applies to iCVD at low Peclet numbers, we measured the residence time distribution (RTD) for argon and isopropanol flowing at 1 sccm, finding a close match to the well-mixed model. We used a Pirani pressure gauge for composition sensing to measure the RTD. We demonstrate the Pirani gauge to be a precise gas composition sensor, which is more modular and less expensive than methods like in-situ infrared spectroscopy and mass spectrometry. We then address heat transfer by measuring thermal diffusion from the filament at a range of pressures in argon and isopropanol. We quantified the effects of non-ideal heat transfer at low pressures and identified a critical Knudsen number, above which heat transfer is independent of pressure. Measurement of thermally driven pressure changes in batch mode enables us to define and model an "effective temperature" for the reactor, matching the thermal diffusion model. Our findings provide a systematic framework for estimating mixing and temperature profiles in iCVD reactors, enabling the assessment of design choices in both batch and continuous operation.

## I. INTRODUCTION

Polymer thin films are used widely in applications such as organic electronics[1] and membranes[2]. Solution phase techniques like spin coating and dip coating are the most popular ways to fabricate polymer thin films, however, the use of polymer Chemical Vapor Deposition (CVD) processes[3] broadens the range of applications. Polymer CVD is solvent-free, enables conformal film synthesis,[4–8] doesn't suffer from surface tension-induced defects like de-wetting, and enables precise thickness control even for ultrathin films (<100 nm).[9] Many polymer CVD processes require harsh conditions (like plasma-enhanced CVD)[10,11] or have a limited library of polymers that can be synthesized (like parylene CVD).[12] Initiated Chemical Vapor Deposition (iCVD) overcomes these limitations by decoupling initiation and propagation. The initiator decomposes at a heated filament, while polymerization occurs on a cooled stage. As a result, iCVD operates under benign conditions and is compatible with a broad range of chemical functionalities and substrates[13].

During iCVD, physical processes like heat transfer from the array and chemical processes like thermal decomposition of the initiator occur simultaneously. In a complex system like this, insights are often gained by using Multiphysics simulations to solve the governing partial differential equations for heat, momentum and mass transfer. We are aware of only two such studies like this in the iCVD literature.[14,15] Both of these studies speculate on novel reactor configurations instead of studying commonly used conditions and, as a result, have not been incorporated into a broader understanding of the iCVD system. Given the importance of Multiphysics simulations to inorganic CVD research,[16] the comparative lack of computational study in iCVD highlights a gap in our understanding.

Though there are few theoretical or computational studies to draw from, elucidation of fundamental processes remains necessary for material design, innovations in synthesis, and optimization for monomer yield[17] and polymer quality. So far, this fundamental understanding comes primarily from studies which model deposition rate, and final film properties (like molecular weight) based on process variables like the reactor pressure and filament temperature.[13] Because these studies do not measure composition in the vapor phase, efforts to unravel the effects of the vapor phase on deposition kinetics have been sparse. While there is an emphasis on understanding the initiator decomposition and surface polymerization reactions occurring in iCVD, these reactions are almost always[18] affected by mixing and heat transfer. Because mixing and heat transfer can be studied in isolation, it is appropriate to establish a strong model of them to set the foundation for addressing reactive processes. While rare, existing efforts to directly measure heat transfer and mixing in the iCVD system warrant discussion.

Regarding heat transfer in iCVD, an excellent experimental investigation of the iCVD filament was performed by Bakker *et al.*, precisely modeling the important radiative effects in heat transfer under variable pressure for vapors relevant to iCVD.[14,19] The model of thermal diffusion proposed by this study is empirical, therefore it doesn't explain how system geometry or gas properties affect heat transfer. While the kinetic theory of gases is expected to explain these phenomena,[20] there have been no efforts to validate a first-principles theory of thermal diffusion in the iCVD reactor.

In the study of mixing, the primary goal is to determine the contacting pattern in the reactor, which generally involves comparing a measured residence time distribution (RTD) with an ideal well-mixed or plug-flow model.[21] The only RTD measurement we are aware of in an iCVD reactor was obtained using the response of vapor composition at the outlet of the reactor to gas pulses measured using an in-situ FTIR.[22] While this work suggests their system is a plug flow, the absence of mathematical analysis and the incompatibility of plug flow with good coating uniformity demonstrated by iCVD under similar conditions cast doubt on the broad applicability of this conclusion.[3] Indeed, the known dependence of mixing on system variables like flowrate[23] suggests that describing flow as either plug flow or well-mixed flow is incomplete. When mixing (or any physical process) depends on system variables, it can be described using dimensionless numbers which correlate with the dominant physical processes driving mixing. While dimensionless numbers have been used in iCVD reactor scale-up,[24] there have been limited efforts to leverage them in the quantitative study of regime changes in iCVD, like the onset of poor mixing or the onset of pressure-dependent heat transfer.[25]

In iCVD, manipulating well-understood regime changes has been remarkably fruitful. For instance, deposition at or above the monomer saturation pressure has yielded insights into areas like polymerization kinetics,[26] nanoparticle fabrication,[27,28] and porous film fabrication.[29] We should expect that less well-understood regime changes in heat and mass transfer might enable similar innovations. For instance, poor precursor mixing in the vapor phase is well known to cause film non-uniformity.[23,24] A formalism to quantify mixing could be used to mitigate the need for batch iCVD,[17,23,30] or be leveraged to produce gradient films (as has been done in parylene CVD).[31] In the realm of heat transfer, it has been shown that heat transfer from the filament is strongly affected by the reactor pressure, but there are no methods of predicting the onset of this effect.[19] The ability to do so might enable the use of initiating systems which are only effective at low pressure.

In this paper, we focus on developing models for both mixing and pressure-dependent heat transfer. The end goal is to allow a researcher to predict the temperature distribution and approximate mixing profile for their reactor using flowrate, pressure and reactor dimensions. To understand mixing in iCVD, precise, real-time knowledge of the vapor phase composition is key. However, vapor phase composition monitoring via in situ FTIR[22] and in situ mass spectrometry[32] are expensive and have not gained widespread usage. Furthermore, they often need to be placed at the outlet of the reactor, potentially introducing a time delay in the measurement. Just as in-situ interferometry enables precise thickness control, an affordable, precise, and simple-to-implement composition sensor opens up possibilities for low-cost composition-based feedback control during deposition. An excellent discussion of the potential benefits of this is contained in Schröder *et al.*.[32]

To address the absence of low-cost, precise and simple real-time vapor phase composition sensing in iCVD, we introduce a method of detecting composition using a Pirani convection gauge. The technique of using the Pirani gauge to obtain real-time vapor-phase composition enables us to compare residence time distribution obtained from simulations to experiment. Pirani gauges operate by measuring heat transferred from a fine wire. At low pressures, the rate at which molecules collide with the wire limits heat transfer, so the rate of heat transfer is proportional to the pressure.[33] Because larger molecules transfer more energy per collision than smaller molecules, the pressure reported by a Pirani gauge depends on the composition of the vapor. In fact, since there are relatively few collisions between gas molecules in a Pirani gauge, the pressure reported by the Pirani gauge often varies linearly with the mole fraction.[34,35] We argue that the potential impacts of Pirani gauge-based composition sensing in iCVD extends beyond what we demonstrate here, pointing to benefits from precise process control and facile deployment in research and the thin film industry.

We begin by estimating dimensionless numbers in iCVD, which suggest an ideal model of heat transfer and mixing in an iCVD reactor in which heat transfer from the filament is pressure-independent and the reactor is well-mixed. Using the model, we show how deviation from the ideal model can be quantified using the Peclet number (mixing) and Knudsen number (Kn) (heat transfer) via pressure-dependent heat transfer experiments and simulations of mixing. While we used argon and isopropyl alcohol (IPA) as model gas species, we demonstrate how our methodology can be generalized to other gases / vapors and reactor geometries. To characterize thermal diffusion, we establish first-principles models which can be generalized to most other reactor geometries to provide predictions of the temperature profile in the reactor. While mixing is more complex, we show that simple, low-cost Multiphysics simulations can be used to evaluate the effect of design choices and estimate the composition throughout the reactor. We hope our work will provide iCVD researchers with the tools to obtain accurate estimates of the mixing state of their reactor, heat transfer in their reactor (including the temperature profile), and how both these phenomena are affected by flowrates, gas composition, pressure, reactor dimensions and reactor temperatures.

## II. EXPERIMENT

### *A. Mixing simulations*

Reactor models used in Multiphysics simulations were built in Autodesk Inventor 2025 (Autodesk, San Francisco California) to match a standard[36] 2-chamber iCVD reactor used in experiments, and a 1-chamber alternative. The 2-chamber design has a top chamber height of 2.25", a bottom chamber height of 3", and a diameter of 9.625". The 1-chamber design has a height of 2.5" and a diameter of 9.625". All dimensions represent dimensions of the vapor volume, i.e. internal dimensions. Schematics for these designs are shown in Fig. 1. Mixing simulations were performed in Autodesk CFD 2024 (Autodesk, San Francisco California). A mixing study was performed with two gases flowing into the reactor through two separate inlets at flowrates of 1 sccm at steady state. The composition at the reactor stage was reported for diffusion coefficients of 4 cm$^2$ s$^{-1}$, 16 cm$^2$ s$^{-1}$, and 64 cm$^2$ s$^{-1}$, corresponding to Pe$_m$ of 5.3, 1.3 and 0.3 respectively at the simulation reactor pressure of 0.1 Torr and 298.15 K. These conditions were chosen to span the upper range of Pe$_m$ for the iCVD system and illustrate the onset of poor mixing.

Transient step-response experiments (Section II.D) were simulated for the 2-chamber reactor under a constant total flow of 1 sccm. To induce the step response, first, a steady, fully developed 1 sccm flow of a primary gas through one inlet was established, then simultaneously a 1 sccm flow of a secondary gas through the other inlet was initiated while the flow of the primary gas was ceased. Step-response curves showing the change in composition were reported for composition at the outlet of the reactor, near the inlet in the top chamber (side port), and far from the inlet in the bottom chamber (back port). Simulations were performed with diffusion coefficients of 4 cm$^2$ s$^{-1}$, 16 cm$^2$ s$^{-1}$, 64 cm$^2$ s$^{-1}$ and 678.5 cm$^2$ s$^{-1}$, corresponding to Pe$_m$ of 5.3, 1.3, 0.3 and 0.03 at the simulation reactor pressure of 0.1 Torr and 298.15 K. The simulation at Pe$_m$ of 0.03 was included to match the conditions for the RTD of IPA and argon measured in Section II.D. Videos of the changing composition are included as supplementary material and follow the naming convention "X_cm2pers.AVI", where X is the diffusion coefficient used.

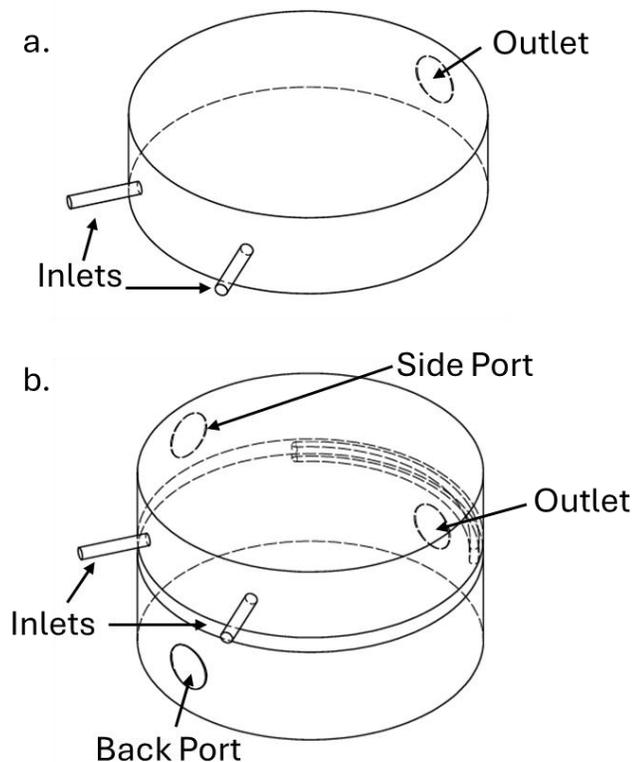

FIG. 1. Schematic of the two iCVD reactor geometries used in simulations (a) single chamber design with inlets and outlet labeled (b) double chamber design with inlets, outlet, side port and back port labeled. This double chamber design matches the design used in experiments.

## *B.   Flow calibration*

Argon gas, obtained from Airgas (Radnor, Pennsylvania) and 99.7% isopropyl alcohol (IPA) obtained from Sigma Aldrich (Burlington, Massachusetts) were used as model gases in all experiments and were both used without further purification. Flow of argon gas was regulated by an MKS mass flow controller type 1159B from MKS instruments (Rochester, New York) while flow of IPA was controlled using an SS-4BMW-VCR needle valve from Swagelok (West Henrietta, New York).

Flowrate measurements used in all experiments for IPA and argon were obtained by flow calibration. First, the leak rate of the reactor was measured by closing the outlet to the pump at base pressure (2-5 mTorr). A gradual increase in pressure due to leak was then monitored using an MKS Baratron capacitance manometer model 626C from MKS instruments (Rochester, New York) by recording pressure every 10 s for 2 minutes and performing a linear regression. The slope of the pressure versus time plot was converted to a flowrate in standard cubic centimeters per second (sccm) using the effective temperature (see Section III.E). The flowrates of gases were measured by setting a flow condition using either the mass flow controller (for argon) or the needle valve (for IPA). The outlet to the reactor was then closed and the increase in pressure was measured every 2 s for 20 s. A linear regression was performed on the pressure versus time plot, and the slope was converted to a molar flowrate in sccm using the ideal gas law. Three replicates were performed for each flowrate.

## C. Vapor phase composition measurements using a Pirani gauge

For binary gas mixtures in the iCVD reactor, the composition-sensitive Pirani gauge pressure is sufficient to deduce composition with good accuracy. First, a standard curve is created by measuring both the Pirani and Baratron pressures at steady state, while varying the vapor-phase composition by changing flow rates of each of the mixing gases (keeping the total flow rate constant). The ratio of the Pirani gauge pressure to the Baratron pressure is then plotted against the vapor-phase composition to create a standard curve (generally linear). In scenarios where the reactor composition is unknown, but the Pirani gauge pressure and Baratron pressures can be read, the standard curve can be used to determine the vapor-phase composition.

## D. Residence time distribution measurement

Step-response curves were generated by flowing argon into the reactor at a fixed flowrate of 1 sccm, then rapidly shutting off the argon flow and opening IPA flow into the reactor such that the total flowrate remained constant. This caused a step change in the composition of inlet vapor, which was registered by a Pirani gauge at the side port. Composition at the side port was determined using a standard curve (Section II.C) matched to IPA/argon mixtures.

## E. Effective temperature measurement

In all experiments, an 80 / 20 Ni Cr nichrome wire (Omega Engineering, Stamford Connecticut) was used for the filament array. The following procedures were used to collect data used to determine the effective temperature in the reactor in Section III. E. These procedures were performed for nominal chamber pressures of 50, 100, 200 and 300 mTorr for both pure argon gas and pure IPA vapor. First, the reactor was pressurized to the nominal chamber pressure. Next, with no power being delivered to the array, the pressure, filament temperature and stage temperature were logged every 5 s for 60 s. Initial leak rate was calculated from the rate of pressure change during this period (detailed in Section II. B). Initial filament temperature was taken as the average filament temperature over the first 60 s period. Subsequently, power to the filament array was stepped up by ~10 W every 60 s until the maximum power, i.e., ~80 W, was reached. Over the course of each 60 s period where power is constant, the pressure, filament temperature, and stage temperature were logged every 5 s.

## F. Heat transfer in argon and IPA vapor

Heat transfer from the filament array to the reactor body can be determined by performing an energy balance on the filament array adapted from Bakker et al.[19] Heat is generated resistively in the filament and is radiated and conducted away from the filament. At low pressures, radiative heat transfer dominates and a radiation-based model which depends only on the filament temperature and stage temperature can be used. At higher pressures, the radiative heat transfer calculated from this model is subtracted from the total dissipated energy to determine the heat transfer by conduction to the gas.

Heat transfer measurements are performed in pure argon gas and pure IPA vapor flowing through the reactor at nominal pressures of 25, 50, 75, 100, 150, 200 and 1000 mTorr and filament temperatures ranging from ~50°C to ~350°C and at constant 1 sccm flowrate of either argon gas or IPA vapor (Section II. B). Power to the filament was measured by multiplying voltage and current from a BK precision DC power supply. Temperature of the filament array was measured using a K-type thermocouple spot-welded to the filament array so that the thermocouple junction was less than 2 mm from the filament to ensure good thermal contact. To verify that current passing through the filament array did not impact the temperature measured by the thermocouple, the current was reversed, and the temperature measurement did not change. Measured values of the filament wire length (2.7 m) and filament diameter (0.4 mm) were used in calculations in Section III. E. Length was determined using total resistance across the array, and the known resistance of the wire at 20ºC of 8.5 Ω m$^{-1}$, while diameter was determined using an instant readout digital caliper from Electron Microscopy Sciences.

# III. RESULTS AND DISCUSSION

## A. *Dimensional analysis*

The first step in modeling heat transfer and mixing in iCVD is to determine which physical phenomena (diffusion, convection, radiation etc.) most significantly impact the state of the reactor. The resulting assumptions simplify both the development and implementation of models. Dimensionless numbers are a well-established method to assess the relative importance of physical phenomena and establish the governing equations for a system. Because the discussion of dimensionless quantities in iCVD literature has been limited to reactor scaleup,[24] further analysis is needed to develop accurate models of heat transfer and mixing.

Table 1. Dimensionless numbers relevant to mass transfer in iCVD

| Name | Equation | Description | Expected Range | Transitional Range |
|---|---|---|---|---|
| Peclet (mass) | $Pe_m = \frac{lv}{D}$ | Advective versus diffusive mass transport | $10^{-6}$-10 | $10^{-1}$-10 |
| Knudsen | $Kn = \frac{\lambda}{l}$ | Length scale of molecular collision to the scale of the system | $10^{-5}$-$10^{-1}$ | $10^{-1}$-10 [37] |
| Reynolds | $Re = \frac{\rho v l}{\mu}$ | Inertial versus viscous forces | $10^{-8}$-1 | 2000-3000[38] |
| Mach | $Ma = \frac{v}{c_s}$ | Characteristic velocity to the speed of sound | $10^{-5}$-$10^{-2}$ | >0.3 [39] |

Dimensionless numbers relevant to mass transfer in the iCVD system are the Reynolds number (Re), Mach number (Ma), Peclet number (Pe$_m$) and Knudsen number (Kn) (summarized in Table 1).[24]

Dimensionless numbers relevant to the transfer of heat include the Peclet number for thermal diffusion ($Pe_t$), the Grashof number (Gr), and the Knudsen number (summarized in Table 2). We do not include dimensionless numbers relevant to the absorption of infrared radiation emitted by the filament because estimating the molar absorptivity of gases is non-trivial and depends both on the emission spectrum of the filament and the absorbance spectrum of the gas. To address the issue of absorbance, we instead compare experiments for absorbing and non-absorbing gases in Section III. D.

Table 2. Table of dimensionless numbers relevant to heat transfer in the iCVD system

| Name | Equation | Description | Expected Range | | Transitional range |
|---|---|---|---|---|---|
| | | | $l = R_{fil}$ | $l = l_{body}$ | |
| Grashof | $G_r = \dfrac{g\rho^2\left(\frac{T_{stage}}{T}-1\right)l^3}{\mu^2}$ | buoyant vs viscous force | | $10^{-5}$-$10^{-1}$ | >$10^2$ |
| Péclet (thermal) | $Pe_t = \dfrac{l v \rho c_p}{\kappa}$ | Conduction versus total heat transfer | $10^{-8}$-$10^{-3}$ | $10^{-6}$-1 | $10^{-1}$-10 |
| Knudsen | $Kn = \dfrac{\lambda}{l}$ | Length scale of molecular collisions to system scale | $10^{-2}$-10 | $10^{-5}$-$10^{-1}$ | >$10^{-1}$ |

Transitional ranges are based on external reference[25], or in the cases of Kn for heat transfer and $Pe_m$ for mass transfer, our own data is presented in further sections. The estimation of dimensionless numbers in Tables 1 and 2 (rounded to the nearest power of 10) were calculated from ranges detailed below which are based on commonly used research conditions.[3]

We have assumed a typical lab-scale reactor with characteristic length-scale ($l_{body}$) ranging from 1 cm (for tubing connections) to 30 cm (long reactor dimension) and a filament radius ($R_{fil}$) of 0.20 ± 0.02 mm. The length scale used to calculate Gr was a typical separation distance between the stage and the filament of 3 cm which does not vary significantly in the literature.[3] The range of reactor conditions are based on common deposition conditions,[13] with stage temperatures ($T_{stage}$) between 0°C and 50°C, filament temperature ($T_{fil}$) from 150°C to 400°C, vapor flowrates from 0.1 to 1 sccm and chamber pressure from 50 to 1000 mTorr. Vapor flowrates and the ideal gas law were used to determine characteristic velocity ($v$) using the cross-sectional area of a typical lab-scale reactor with approximate height of 5 cm and long-reactor dimension (width or diameter) of 30 cm. In calculation of Gr, the acceleration due to gravity was 9.81 m s$^{-2}$.

Gas density ($\rho$) and heat capacity ($c_p$) were determined for ideal gases with 3 rotational and 3 translational degrees of freedom (heat capacity ratio, $\gamma$, of 4/3). Diffusivity ($D$), dynamic viscosity ($\mu$), mean free path ($\lambda$) and speed of sound ($c_s$) were calculated based on correlations found in Poling[40] and Lide[41] for either a "large" or "small" process gas molecules. The following gas properties used in estimations were based on tabulated data for molecules similar to those used in iCVD.[40,41] The thermal conductivity ($\kappa$) was estimated directly from tabulated data[42] and the assumption that $\kappa \sim \sqrt{T}$ from kinetic theory of gases[41] to range from 0.001 $T^{1/2}$ to 0.0005 $T^{1/2}$ W m$^{-1}$ K$^{-1}$ for "small" and "large" cases respectively. Molecular weight (M) of process gases was assumed to range from 50 to 500 g mol$^{-1}$,[13] boiling point of process gases from 200 to 50°C (based on values reported by suppliers for common monomers) and critical molar volume ($V_c$) of process gases between 200 and 1000 cm$^3$ mol$^{-1}$ for "small" and "large" cases respectively.[41] The temperature ($T$) used to determine gas properties was assumed to range from the minimum stage temperature (0°C) to the maximum filament temperature (400°C).

For calculation of mean free path, viscosity and diffusivity, the collision diameter ($d$) of process gases was calculated using Eq. 9-48 from Poling.[40] In this correlation the method of Tyn and Calus was used to calculate the molar volume of process gases at their boiling point as reported in Lide.[41] The mean free path was calculated from kinetic theory using Eq. (1),[41]

$$\lambda = \frac{k_B T}{\sqrt{2}\pi d^2 p} \quad (1)$$

where $p$ is the chamber pressure and $k_B$ is Boltzmann's constant. Diffusivity and dynamic viscosity of process gases was calculated using the method of Wilke and Lee as reported in Poling.[40] The speed of sound was calculated using the relationship for ideal gases given by Eq. (2),[43]

$$c_s = \sqrt{\frac{\gamma R_G T}{M}} \quad (2)$$

where $R_G$ is the ideal gas constant.

Re is the ratio of inertial to viscous forces in a flow and is often used to establish the onset of turbulence.[38] Because vapors in iCVD are low density, viscous stresses dominate inertial stresses under typical conditions, so Re is small, and flow is laminar.[24] Ma relates the speed of sound to the characteristic velocity of the flow and is often used to determine whether a flow is incompressible.[39] The typical range of volumetric flowrates used in iCVD (discussed above) results in low characteristic velocities, meaning Ma is much less than its transitional value[39] of 0.3, and compressibility effects can be neglected. Because Ma << 0.3 and Re << 2000,[38] we determined that typical vapor flow in iCVD is laminar and incompressible. Furthermore, because Re << 1 in most cases, the flow profile in the reactor is well-described by equations for Stokes flow.[38]

Both $Pe_m$ and Kn are of a transitional scale, meaning that, while the effects they describe can often be neglected, they become important under certain conditions. Kn is the ratio of the mean free path to the length scale of the system. When Kn is less than ~0.1, the flow is 'viscous' and can be described by continuum fluid mechanics.[37] Kn is anticipated to be of transitional scale in confined parts of the reactor (such as tube connections) when small process gases are used at low pressures. Under these conditions, flow transitions to Knudsen flow, a regime between viscous and molecular flow, where continuum assumptions break down.[37] Because our study focuses on the bulk reactor and Knudsen flows primarily occur in tube connections, we will not investigate Knudsen flows further. It should be noted that Knudsen flow can be important for reactor design, since it can play a role in long pump-down times when long, small-diameter tubing connections are used.

$Pe_m$ describes the ratio of advection to diffusion. When $Pe_m$ is small (for low flowrates or high diffusivity mixtures), the well-mixed assumption is valid. Because well-mixed conditions correspond with uniform coating in iCVD,[30] the well-mixed condition is the ideal mixing condition. As $Pe_m$ increases, its value quantifies the departure from perfect mixing. Because the expected range of $Pe_m$ for the iCVD system overlaps with the transitional range[25] of $Pe_m$ (Table 1), we will investigate it further in Sections III. B and C.

Table 2 summarizes dimensionless numbers relevant to heat transfer in iCVD. It should be noted that we do not include dimensionless numbers related to radiation in this discussion. While thermal radiation makes up a significant fraction of the heat emitted by the filament,[19] our experiments in Sections III. D and E provide no evidence that radiation impacts process gases. Because heat flux is inversely proportional to distance from a slender body like the heated filament in iCVD,[44] and most of the conductive resistance to heat transfer occurs close to the filament surface (see Section III. D for a more detailed discussion). , $Pe_t$ for heat transfer from the filament is small, and heat transfer is predominantly conductive. Because $Pe_t$ is small on the scale of the filament array (when the characteristic length scale is the filament radius) its role in heat transfer is expected to be small, and we will leave any investigation of thermal advection in iCVD for future work.

Kn is defined the same for heat transfer as it is for mass transfer. In heat transfer, our primary interest in Kn is over the length scale of the filament array. Kn based on filament length scales describes the effect of low pressure on heat transfer near the filament. Because the conductive heat flux across most surfaces in the reactor is small, the role of Kn is only significant at the filament array.

Gr describes the ratio of buoyant to viscous forces in a fluid. When Gr is above $10^2$, natural convection begins to occur. Though there has been speculation about the role of natural convection in iCVD,[15] for the typical conditions in Table 2, we find that the Grashof number is three orders of magnitude below the range at which natural convection is known to occur. As such, the description of heat transfer from the walls of the reactor can be described primarily by Fourier conduction. Our discussion of dimensionless numbers in heat transfer does not seek to be fully comprehensive over all potential iCVD conditions but accounts for the most significant impacts on heat transfer to the main process gas for typical operating conditions.

## B. *Mixing simulations*

Simulations were performed to help visualize the mixing between two distinct gas flows entering the reactor through separate inlets. Two reactor geometries were used. The double chamber geometry matched the reactor used in experiments, and a single chamber equivalent was simulated to show the effects of design modification. While 2-chamber designs are common, the bottom chamber acts as a difficult-to-model dead zone. Steady-state mixing simulations were performed with a constant flow of gas into the reactor through both inlets at 1 sccm. To obtain models of the full range of $Pe_m$, we varied the diffusion coefficient. Simulations were performed for $Pe_m$ of 0.3, 1.3 and 5.3 in both the 1-chamber and 2-chamber designs, indicating a transitional $Pe_m$ of roughly 0.1-10. Simulations were also performed for $Pe_m$ of 0.03, roughly corresponding to the mixing of argon and IPA studied in Section III.C. Under these conditions, the system was well-mixed, corresponding to the ideal case in Fig. 3 (details can be found in the supplementary material). Because poor mixing would be associated with non-uniform and difficult to repeat or characterize iCVD deposition, these simulations indicate that for good performance, iCVD reactors should be operated below $Pe_m$ of 0.3. $Pe_m$ for IPA / argon mixing system was calculated based on the binary diffusion coefficient of IPA and argon reported in Lee[45], scaled down to a pressure of 100 mTorr using the pressure dependence suggested by kinetic theory of gases[41]. We have calculated $Pe_m$ from the individual gas flowrates, not the overall gas flowrates such that each gas flow is associated with a different value of $Pe_m$.

It is notable from simulations (Fig. 2) that the increased volume from the 2-chamber design has very little effect on the transitional $Pe_m$; as such, the appropriate length scale to use in calculating $Pe_m$ is the distance from the inlet to the outlet. The minor difference in the composition profile of the 1-chamber versus 2-chamber design near the outlet of the reactor is caused by the slit shaped connection between the top and bottom chambers for the 2-chamber design (Fig. 1). While we only show simulations of circular "pancake" type reactors, if $Pe_m$ for a more exotic reactor geometry needs to be calculated, the longest dimension should be used as the length scale. This is because during actual deposition, not only do inlet streams need to be well mixed, but degradation products of the pre-initiator must also be well-mixed (i.e. poor mixing and plug flow are both non-ideal mixing scenarios). It should be noted that $Pe_m$ describes non-ideal mixing; however, even at low $Pe_m$, non-uniform depositions can occur at high deposition rates[29].

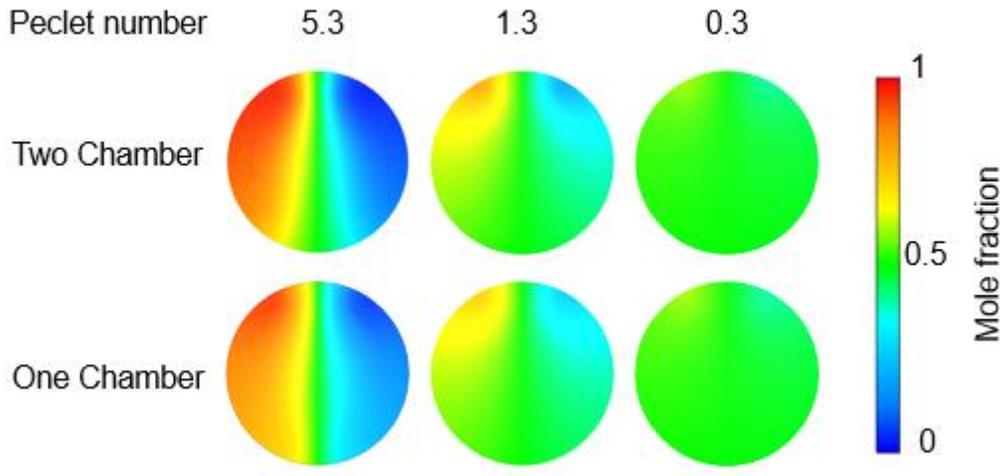

FIG. 2. Binary gas mixing for a range of $Pe_m$ over the transitional range of $Pe_m$ in two-chamber and one-chamber reactor configurations.

Further transient simulations were performed to provide insight into the step response experiments performed in Section III.C. While impulse response curves are generally preferred to step response in determining contacting patterns and mixing,[21] generating and detecting an impulse without significant pressure-driven flows was infeasible in our setup. Simulations were run to exactly replicate the procedures used in Section III.C. Fig. 3 shows step response for all 3 $Pe_m$ of 5.3 (Fig. 3 (a)), 1.3 (Fig. 3 (b)) and 0.3 (Fig. 3 (c)). Simulations were also performed for a $Pe_m$ of 0.03, roughly corresponding to the mixing of argon and IPA studied in Section III.C, and this curve could not be distinguished from the theoretical ideal. Videos of the simulations can be found in the supplementary material as well as data for $Pe_m = 0.03$.

The step response curves in Fig. 3 enable two important conclusions. First, for $Pe_m$ associated with poor mixing in Fig. 2, significant deviation from ideal mixing response was observed via the side port and backport, demonstrating that step-response experiments (Section III. C) could assess mixing for a given set of conditions. Second, the data demonstrates clearly the sensitivity to sensor location. In practice, relying on the shape of the step response curve can introduce significant errors, since generating the step change can induce moderate pressure-driven flow in the system, which is not easy to decouple from time delay at high $Pe_m$. Furthermore, precise step response data require accurate flowrate calibration, which tends to be subject to experimental error when working with low diffusivity gases. Even for an extremely poor mixing scenario (Fig. 3 (a)), it may be difficult to confirm poor mixing if the Pirani gauge is placed on the outlet of the reactor, so this should be avoided. Because gas sensing techniques like in-line FTIR often analyze the exhaust gas from the reactor[22], they are much less valuable for real-time sensing of reactor dynamics. Furthermore, at a $Pe_m$ of 1.3 (Fig. 3 (b)) it may become challenging to decouple pressure shock from non-ideal behavior, even if sampling at the back port or side port is performed. The simulations in Fig. 3 (c) show that if precise mixing information is needed, sensors should be placed in multiple locations far apart in the reactor, and their difference or ratio should be used to provide a sensitive measurement of deviation from ideal mixing. Step response curves in single chamber reactor configurations may differ from Fig. 3. However, the correspondence between 1-chamber and 2-chamber designs in Fig. 2 suggests the insights from Fig. 3 on sensor placement apply to other reactor geometries.

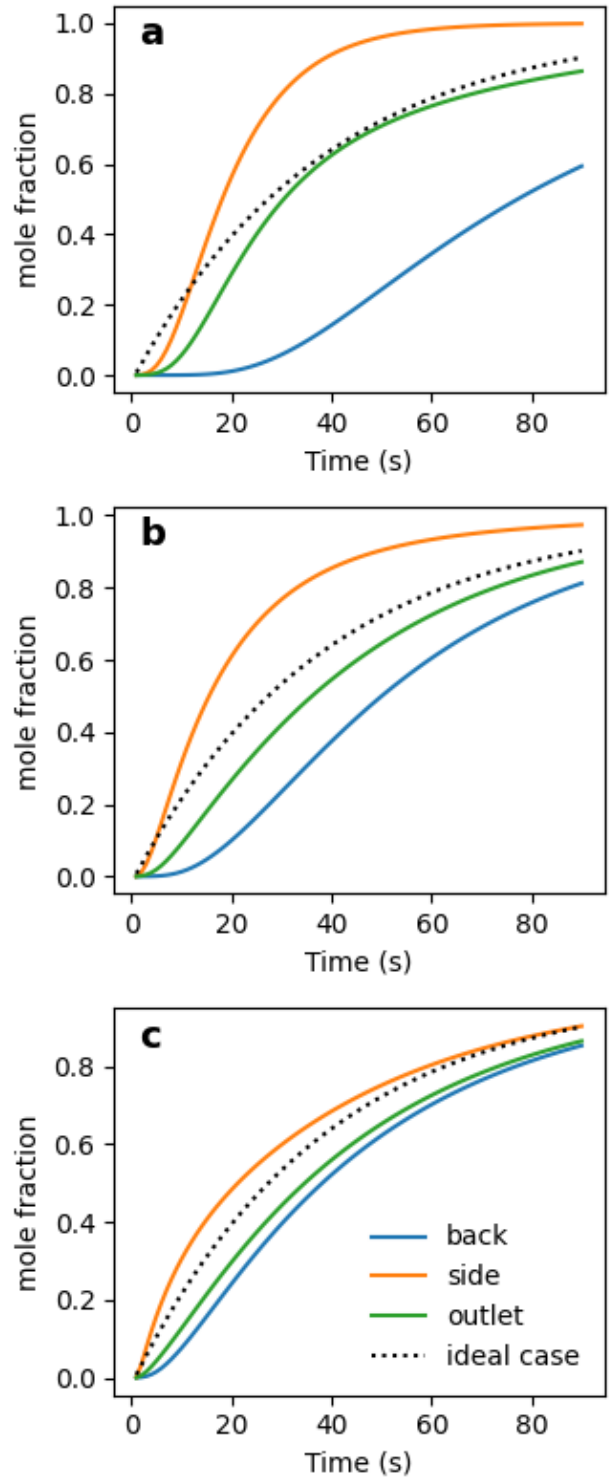

FIG. 3. Step response curves for 2-chamber reactor configuration comparing ideal mixing to various mixing scenarios simulated using different $Pe_m$. The mole fraction of a gas species introduced as a step function into the chamber ($y$) is sampled at three locations in the simulation: the back port of the reactor, the side port, and the reactor outlet. Simulations are presented for (a) $Pe_m$ of 5.3, (b) $Pe_m$ of 1.3 and (c) $Pe_m$ of 0.3.

## C. Step response measurements

In-situ composition sensing has been used on rare occasions in the iCVD literature,[22,32] However, a rigorous study of contacting patterns in iCVD and the phenomena responsible for them has been absent. In contrast to costly custom systems for composition sensing,[22,32] we present a simple, off-the-shelf method of determining binary composition data using a convection-assisted Pirani pressure gauge. The Pirani gauge uses the pressure dependence of heat transfer to determine pressure and provides a precise, but composition-dependent, measurement of pressure in the range of 1 mTorr to 1 Torr,[34] which corresponds to the pressure range for iCVD.[13] Because heat transfer in low-pressure gases depends on the composition of a gas, Pirani gauges are not appropriate for determining pressure in an iCVD system; however, when combined with a Baratron capacitance manometer, the Pirani gauge becomes an effective composition sensor. Thermal conductivity is routinely used for gas composition sensing[46], so the use of the Pirani gauge is not a significant departure from existing methods. However, it has never been used in the iCVD literature, so our methods represent an exciting new capability for monitoring and controlling the reactor environment. To measure the composition, a standard curve is first created (Fig. 4 (a)) in which IPA vapor and argon gas are flowed into the iCVD chamber to obtain a range of compositions, the pressure reported by the Pirani gauge is then recorded and divided by the pressure reported by the Baratron capacitance manometer to yield a proxy composition measurement we will refer to as the Pirani gauge ratio (*c*). The resulting standard curve is linear and can be used to convert the Pirani gauge ratio to composition.

Real-time composition sensing from the Pirani gauge enables step-response experiments in which argon is initially flowed at a steady-state, then it is quickly shut off and replaced with a flow of IPA vapor. This experiment maintains a constant total flowrate and pressure. The resulting step response curve (Fig. 4 (b)) shows a characteristic well-mixed response (often called a continuous stirred tank reactor or CSTR model). As such, the well-mixed assumption describes the contacting pattern for the reactor conditions and flow rates studied. The apparent presence of bypass in the residence time distribution (Fig. 4 (c)) is indicative of a slight pressure-driven flow when the inlet composition is changed. While a pulsed response is generally more sensitive to non-ideal reactor conditions[21], we use a step-response to maximize the sensing resolution, minimize the effect of pressure-driven flows, and importantly, to maintain the integrity of the signal – it is significantly easier to introduce a step change at constant flowrate and pressure than it is to introduce a pulse which approximates a delta function to an iCVD reactor system.

Based on the measured flowrate of 1.04 sccm into the reactor, the step-response curve indicates a reactor volume of 5.7 L, which is close to the expected reactor volume measured of 6.3 L (see supplementary material for calculation). These results are not surprising, given the expectation that the conditions measured represent a $Pe_m$ of 0.03 (based on reactor dimensions and a calculated diffusion coefficient of 678.5 $cm^2/s$).[45] Furthermore, our earlier simulations indicated that the reactor would behave as a well-mixed system at these values of $Pe_m$ (Fig. 2 and 3). Nevertheless, the realization of a well-mixed vapor phase is in marked contrast to the current literature,[22] which has concluded that the iCVD reactor system is a plug flow system. We believe the contrast may be due in part to the pulse response curves used in prior work,[22] which are associated with the aforementioned drawbacks. Past simulations of iCVD reactors, which show poor mixing, have generally utilized reactor geometries and conditions that differ significantly from standard research conditions.[14,15]

It is clear from dimensional analysis (Section III. A) that the iCVD reactor mixing system is determined by $Pe_m$ and Kn, with Re and Ma indicating that the flow is decisively laminar and incompressible. The measured step response curves for IPA and argon illustrate that the reactor system is well-mixed, in agreement with simulation results in Section III. B. Together with Sections III. A and B, the data in Fig. 4 supports the use of the well-mixed assumption for the iCVD reactor at low $Pe_m$, with reduced coating thickness/composition uniformity as $Pe_m$ increases.

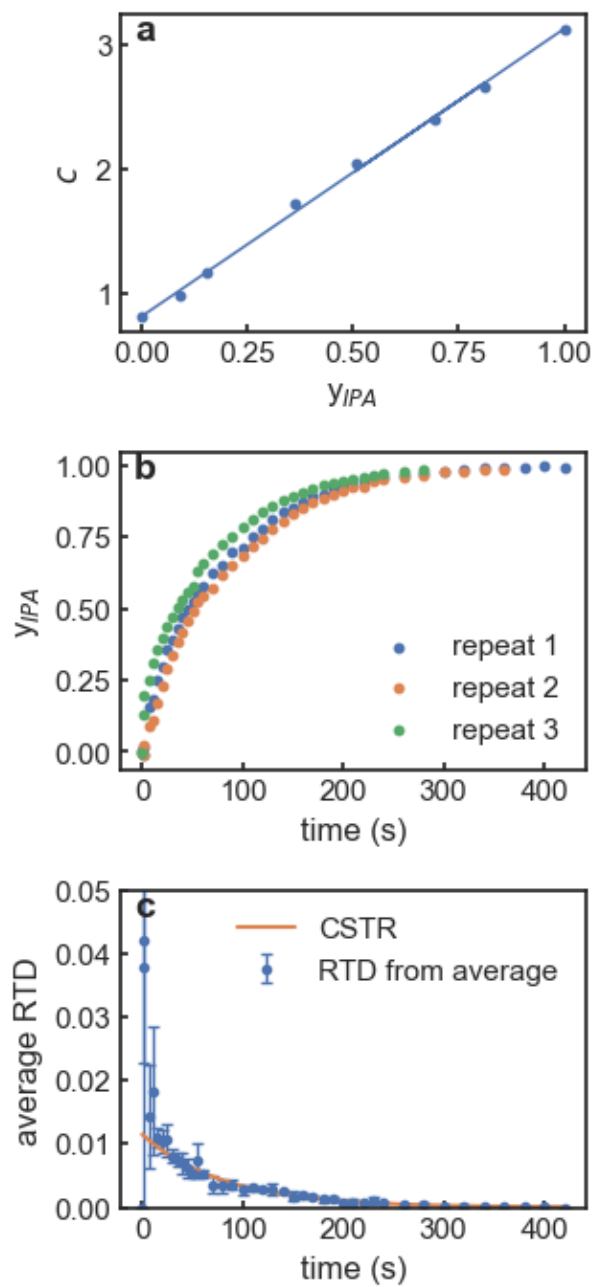

FIG. 4. Measured step-response and residence time for the IPA and argon mixing system. (a) Standard curve correlating the Pirani gauge ratio ($c$) to the mole fraction of IPA in a vapor phase made of IPA and argon. (b) Experimentally measured step response curves for the mole fraction of IPA versus time (IPA was introduced at $t = 0$), calculated from the IPA/argon standard curve in (a). (c) The average residence time distribution (RTD) for IPA, derived from the step response curves shown in (b), which fits well to a well-mixed ideal.

## D. Heat transfer

The work of Bakker et al.[19] established the measurement of heat transfer from the filament wires in iCVD, developing a radiation model to describe heat transfer at low pressure. Empirical models for the diffusive heat transfer from the filament were also introduced by Bakker et al. However, the derivation of first-principles models of diffusive heat transfer is necessary to predict heat transfer in novel reactor geometries or vapors. We perform experiments similar to Bakker et al. in which we measure filament temperature while varying the pressure in the reactor and the power delivered to the filament array. Following from Bakker et al.,[19] heat transfer by conduction from the filament array in W m$^{-1}$ ($q_{fil}$) is given by Eq. (3) in accordance with radiation between grey bodies, when the visible area of one (the walls of the reactor) is much larger than the other (the filament)[47],

$$q_{fil} = \frac{RI^2}{L_{fil}} - 2\pi R_{fil} \sigma \epsilon_{fil} \left(T_{fil}^4 - T_{stage}^4\right) \quad (3)$$

where $R$ is the resistance of the nichrome filament wire, $I$ is the current through the filament, $L_{fil}$ is the length of the filament wire, $\sigma$ is the Stephan Boltzmann constant, $R_{fil}$ is the radius of the filament wire and $\epsilon_{fil}$ is the emissivity of oxidized nichrome wire, which has a value of 0.95-0.98.[47]

To model diffusive heat transfer in the iCVD vapor, we first consider that diffusive heat flux in W m$^{-2}$ ($\boldsymbol{j_q}$) follows Fourier's law given by Eq. (4),

$$\boldsymbol{j_q} = -\kappa \nabla T \quad (4)$$

where $\kappa$ is the thermal conductivity. At steady-state, conservation of energy is given by Eq. (5).

$$\nabla \cdot \boldsymbol{j_q} = 0 \quad (5)$$

According to the kinetic theory of gases[33] $\kappa \sim \sqrt{T}$, therefore, we can define a heat transfer potential $\phi$ from Eq. (4) such that $\boldsymbol{j_q} = \nabla \phi$ to rewrite Eq. (5) as Eq. (6),

$$\nabla^2 \phi = 0 \quad (6)$$

where $\phi$ is defined by Eq. (7),

$$\phi = -\frac{2}{3}\kappa' T^{\frac{3}{2}} \quad (7)$$

where $\kappa'$ is the temperature independent thermal conductivity given by Eq. (8).

$$\kappa' = \frac{\kappa}{\sqrt{T}} \qquad (8)$$

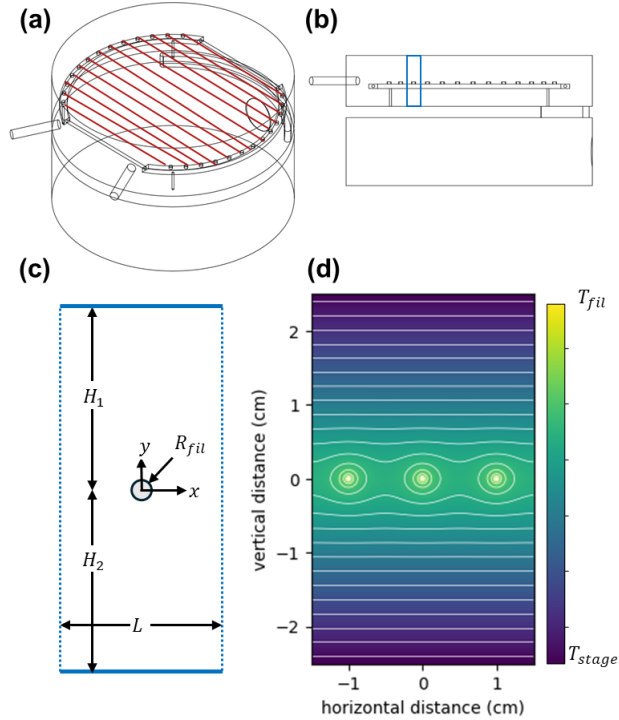

FIG. 5. Schematic of filament array placement and heating. (a) Oblique view of the 2-chamber reactor showing the placement of the filament array in the top chamber. (b) Side view of the reactor with the blue box corresponding to the schematic in (c). (c) Schematic showing the coordinate system used in the modeling of temperature at the filament array (d) Plot showing Eq. (9) for 3 adjacent filament wires as a visual aid.

To obtain a model for heat transfer in the reactor from the filament (Fig. 5 (a)), we must now solve Eq. (6) for the reactor vapor phase. We will simplify the reactor geometry by ignoring the edges of the filament array so we can treat it as a 2D problem given by the domain shown in Fig. 5 (b) and (c), where we have a boundary condition at the stage and lid ($y = -H_2$ and $H_1$) that $\phi = \phi_{stage}$ and $\phi_{lid}$ respectively and the periodic boundary condition that $\boldsymbol{j_q} = 0$ at $x = -L/2$ and $L/2$, where $x$ is horizontal distance from the filament, $y$ is vertical distance from the filament, $H_1$ and $H_2$ are the distances from the filament to the lid (22.86±0.02 mm) and stage respectively (31.75±.02 mm), and $L$ is the distance between filaments (14.7±0.2 mm) as shown in Fig. 5 (c). Next, we note that the filament wires are close to each other relative to the height of the chamber, meaning that the solution of Eq. (6) for an infinite line of point sources closely approximates the boundary conditions in Fig. 5 (c). This assumption is consistent with the design of CVD reactors with heated filaments, because sufficiently large filament spacing to cause non-uniform stage heating also causes non-uniformity in the film.[48]

To find the solution of Eq. (6) for an infinite line of sources in two dimensions, we can draw from the field of complex analysis, in which the Cauchy-Reimann equations show that the real and imaginary parts of any complex function are solutions to Laplace's equation (Eq. (6)).[49] Therefore, obtaining the solution for Eq. (6) for an infinite line of equally spaced 2D point sources (line sources), consists of finding a complex function with an infinite line of equally spaced singularities. Just as $\text{Re}(\ln(z))$ corresponds to a single point source in $x,y$-coordinates when $z = x+iy$, it can be seen that $\text{Re}(\ln(\sin(z)))$ corresponds to an infinite line of point sources (elaboration on this can be found in the supplementary material). When complex numbers are converted to cartesian coordinates, and constants of integration are determined from the boundary conditions discussed above, we obtain Eq. (9) (see supplementary material for details of the derivation),

$$\phi = \phi_R + \frac{q_{fil}}{2\pi} \ln\left(\frac{L}{R_{fil}\pi} \sqrt{\sin^2\left(\frac{x\pi}{L}\right)\cosh^2\left(\frac{y\pi}{L}\right) + \cos^2\left(\frac{x\pi}{L}\right)\sinh^2\left(\frac{y\pi}{L}\right)}\right)$$
$$+ \left(\frac{\phi_{lid} - \phi_{stage}}{H_1 + H_2} - \frac{q_{fil}}{2L}\frac{H_1 - H_2}{H_1 + H_2}\right)y$$
(9)

where $\phi_R$ is the values of $\phi$ at the surface of the filament, analogous to $T_e$ in Su *et al.* (distinct from the heat transfer potential of the filament itself $\phi_{fil}$). To help visualize the temperature distribution in the reactor, Fig. 5 (d) shows an example plot of Eq. (9) for 3 adjacent filament wires.

To compare Eq. (9) to experimental data, we first assume the measured temperature of the stage is approximately equal to the average of the stage and lid temperatures. Justifying this assumption, our data shows that changes in stage temperature during experiments affect the overall driving force ($\phi_{stage} - \phi_{fil}$) by ~ 5%, meaning that slight differences in $T_{stage}$ and $T_{lid}$ have a minor effect on model validity. Proceeding with this assumption, we find that the overall heat transfer resistance caused by the bulk gas can be described by Eq. (10).

$$\frac{(\phi_{stage} - \phi_R)}{q_{fil}} = \frac{1}{2\pi} \ln\left(\frac{L}{2R_{fil}\pi}\right) + \left(\frac{H_1 H_2}{H_1 + H_2}\right)\frac{1}{L} \quad (10)$$

To relate this model to the filament temperature (which we measure), we propose an interfacial heat transfer resistance in accordance with models of the Knudsen jump in planar geometry.[20]

$$\phi_{fil} - \phi_R = -\frac{b\lambda q_{fil}}{2\pi R_{fil}} \quad (11)$$

The mean free path is calculated from Eq (1) using collision diameters of 0.58 nm and 0.358 nm[41] for IPA and argon respectively. The diameter of IPA was approximated using equation 9-4.8 in Poling[40] and a critical volume of 222 cm$^3$/mol[41]. The dimensionless interfacial heat transfer resistance $b$ is approximately described by Eq. (12) according to the kinetic theory of gases, and the Boltzmann theory of interfacial heat transfer,[20,50]

$$b = \frac{8}{9}\frac{2-\alpha_0}{\alpha_0(\gamma+1)} \quad (12)$$

where $\alpha_0$ is the accommodation coefficient or the fraction of inelastic collisions with the filament. Note that in fitting experimental data, the heat capacity ratio was 1.105 for IPA[51] and 1.66 for argon (the value for a monatomic ideal gas). Explanation of how Eqs. (11) and (12) are extracted from Su et al.[20] can be found in the supplementary material. Combining Eqs. (10) and (11), to give Eq. (13) represents the overall heat transfer from the lid to the stage.

$$\phi_{fil} = \phi_{stage} - \frac{q_{fil}}{2\pi}\left(\ln\left(\frac{L}{2R_{fil}\pi}\right) + \left(\frac{H_1 H_2}{H_1 + H_2}\right)\frac{2\pi}{L} + \frac{b\lambda}{R_{fil}}\right) \quad (13)$$

Further details on the derivation of Eq. (13) can be found in the supplementary material.

Data was collected and processed according to the following procedures. At low pressure, $q_{fil}$ (heat transfer by conduction from the filament array) becomes negligible, so we can solve for the wire emissivity from this limit.[19] To do this, the filament temperature was measured at base pressure (~3 mTorr) while varying power to the filament. Temperature versus power data was then fitted to Eq. (3), by varying the wire emissivity, assuming $q_{fil}$ to be zero. This fit is shown in Fig. 6 (a) and corresponds to an emissivity of 2.14, which is notably unphysical (emissivity must be less than 1). It is likely that this high emissivity value is caused by a small leakage of current through the feedthroughs to the reactor, resulting in a mismatch between the current reading from the power supply and the actual current passing through the filament array. Other factors, like error in the length of the filament or conductive losses, are not large enough to cause such a discrepancy. In all calculations, the current is adjusted by a constant factor such that the emissivity of the array is 0.95.

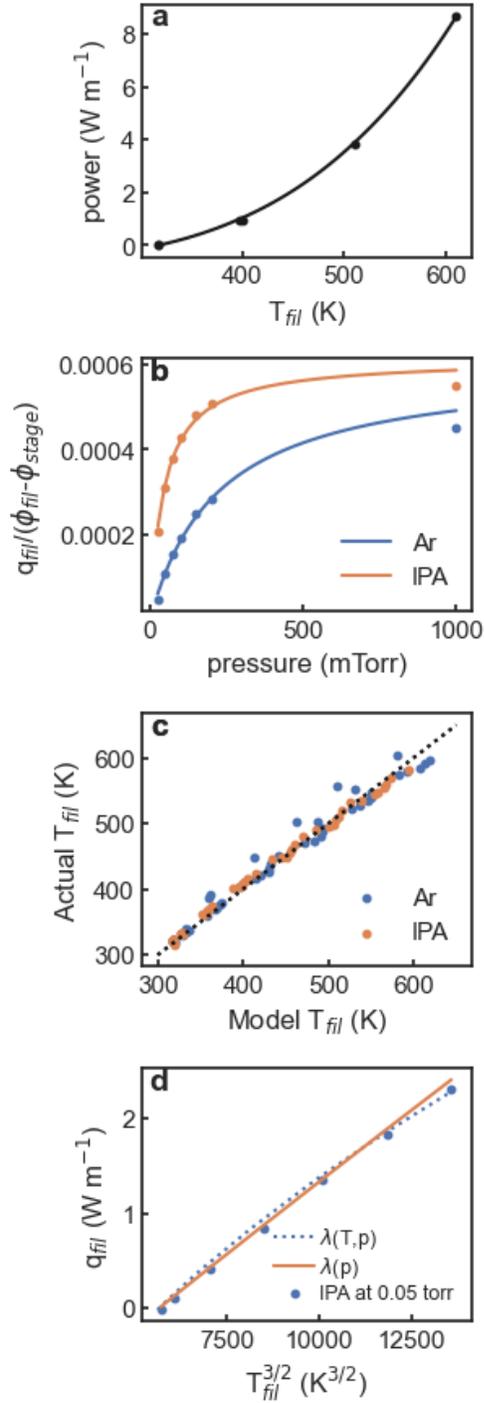

FIG.6. Experimental data and curve fits of heat transfer from the filament array to the reactor body. (a) Power versus filament temperature in the low-pressure limit (3 mTorr) including a curve fit of Eq. (3). (b) Overall thermal conductivity versus pressure used to obtain initial values for the fitting parameters in Eq. (13), where temperature dependence

of mean free path is neglected. (c) Actual filament temperature (blue and orange dots) versus predicted filament temperature (dotted line) based on Eq. (13) for IPA (orange) and argon (blue). (d) Comparison between preliminary curve fit and final curve fit for conducted heat flow versus filament temperature, for Eq. (13) when we account for temperature dependence of the mean free path (dotted line) and ignoring the temperature dependence of the mean free path (solid line).

From the known power to the array and radiative heat loss, $q_{fil}$ can be determined from Eq. (3). By collecting power and filament temperature data at a range of chamber pressures (up to 1 Torr) and calculating $q_{fil}$, we can then fit Eq. (13) to experimental data. Before fitting Eq. (13), however, initial guesses for the fitting coefficients, $\kappa'$ and $\alpha_0$, were determined. This was done by ignoring the temperature dependence of the mean free path (calculating it at a fixed 450 K). For each chamber pressure, this allowed the proportionality constant relating $q_{fil}$ to the thermal driving force ($\phi_{fil} - \phi_{stage}$) to be calculated and fit to Eq. (13), to obtain initial values for the fitting parameters $\kappa'$ and $\alpha_0$ (Fig. 6 (b)). This proportionality constant can be thought of as overall thermal conductivity, with Fig. 6 (b) clearly showing that the reactor vapor phase becomes more insulating as pressure drops for both gases. Note that because of heat losses, filament temperatures were lower at higher pressures, so the nominal 450 K for mean-free path estimation leads to an apparent over-prediction of the filament temperature for the data taken at 1 Torr. Fig. 6 (b) also clearly shows that there is little pressure dependence above pressures of roughly 400 mTorr and 1 Torr for IPA and argon respectively. These correspond to a critical Kn of 0.4-0.45. Assuming the same critical Kn, we expect to see very little pressure dependence for monomer and initiator molecules commonly used in iCVD at normal operating pressures.

Using the initial values from the fit shown in Fig. 6 (b) the fit was repeated (except this time, the mean free path (Eq. (1)) was evaluated at the filament temperature). Data was fit to Eq. (13), by varying $\kappa'$ and $\alpha_0$ to minimize the mean squared error between the predicted and measured values of the filament temperature. The fitting data encompasses the range of pressures and filament temperatures commonly used in iCVD for both IPA vapor and argon gas. The resulting predicted versus measured $T_{fil}$ values are plotted in Fig. 6 (c). Fitting parameters from Eq. (13) are included in Table 3. To show how the overall thermal conductivity plotted in Fig. 6 (b) compares to the fitting results shown in Fig. 6 (c), $q_{fil}$ is plotted against filament temperature with the fits based on temperature-independent and temperature-dependent mean free path, respectively, for IPA at 50 mTorr (Fig. 6 (d)). This comparison shows little impact from the temperature dependence of the mean free path.

TABLE 3. Parameters used to fit Eq. (13)

|     | $\kappa'$ (W m$^{-1}$ K$^{-3/2}$) | $\alpha_0$ (non-dim) |
|-----|---|---|
| IPA | 0.00119 | 0.33 |
| Ar  | 0.00117 | 0.15 |

The thermal conductivity of IPA from Lange's Handbook[42] is 0.025 W m$^{-1}$ K$^{-1}$ at 250°C, giving a $\kappa'$ of 0.0011 W m$^{-1}$ K$^{-3/2}$, while argon is reported to have a thermal conductivity of 0.0211 W m$^{-1}$ K$^{-1}$ at 100°C, giving a $\kappa'$ of 0.00109 W m$^{-1}$ K$^{-3/2}$. The values of $\kappa'$ reported in Table 3 match well with the literature values. They vary from the reported values by the same constant, which is likely due to slight inaccuracy in the system parameters or edge effects not accounted for in Eq. (13). The accommodation coefficient $\alpha_0$ describes the impact of the Knudsen jump based on a calculation of the mean free path from Eq. (1). The values of $\alpha_0$ show that, when correcting for the molecular diameter in the mean free path, the argon has a less efficient thermal accommodation[20] at the surface of the filament. It is important to note that a planar model was used to approximate the filament array in describing the relationship between the Knudsen jump and the accommodation coefficient, which may not apply to the cylindrical geometry of the actual filaments and could contribute to the low $\alpha_0$ values. Furthermore, the outliers in Fig. 6 (c) (the ones not well predicted by the heat transfer model) correspond to argon gas at low pressure. Under these conditions, the assumption of Eq. (11), corresponding to a planar Knudsen jump, is less appropriate, and geometry-dependent corrections[52] can be introduced if necessary.

Our results for heat transfer from the filament in the iCVD system show that the pressure dependence of conduction can be well accounted for by models of the heat transfer. Furthermore, they show that the same conduction model is valid for gases which absorb infrared radiation (IPA) and those which do not (argon). As such, the predictive ability of our model experimentally validates that infrared absorption can be neglected at iCVD conditions, as was discussed in Section III.A. We expect our results to be helpful in elucidating the thermal decomposition of di-*tert*-butyl peroxide at the filament array, which has largely gone unexplained. The model we propose corresponds well with the existing theory of heat transfer and reported heat transfer coefficients (Table 3), suggesting it is a strong descriptor of heat transfer in the iCVD reactor system. However, due to the neglect of edge effects in our description of heat transfer, it is still an approximate theory, and further work could improve the precision in the estimation of heat transfer if needed.

### E.    Reactor effective temperature

When operating the iCVD reactor in batch mode,[17,27,28,30] the ideal gas law is a helpful model. In fact, most studies of iCVD implicitly use the ideal gas law to perform flowrate calibration before running a deposition. Whenever the ideal gas law is used in these situations, the temperature of the reactor vapor must be specified, however, because the temperature in the reactor is non-uniform, it's not immediately clear what the temperature of the reactor is. The effect of temperature gradients becomes even more pronounced in any batch mode reaction or deposition, such as those that are used to perform condensed droplet polymerization.[27,28] As batch processes become more common in the literature,[17,30] the analysis and control of these processes will require a precisely-defined temperature. Herein, we refer to the temperature consistent with treating the reactor with the ideal gas law as the effective temperature ($T_{eff}$). By turning on the filament array while the reactor is closed off from the pump, we measured the change in the effective temperature caused by the filament array.

To define the effective temperature, we begin by assuming (based on Section III.A) that compositional and pressure gradients in the iCVD reactor are small, but we acknowledge that due to thermal gradients, vapor density $\rho$ varies with location $x$ such that at any location the ideal gas law is valid as described by Eq. (14).

$$T(x) = \frac{p}{R_G \rho(x)} \quad (14)$$

Integrating Eq. (14) over the volume of the reactor, we obtain Eq. (15),

$$\int \frac{1}{T} dV_r = \frac{nR_G}{p} \quad (15)$$

where $n$ is the total number of moles of gas in the reactor. To appeal to the ideal gas law for the reactor overall, we define the effective temperature of the reactor such that the ideal gas law holds for the whole reactor ($pV_r = nR_G T_{eff}$), giving Eq. (16),

$$T_{eff} = \frac{V_r}{\int \frac{1}{T} dV_r} \quad (16)$$

where $V_r$ is the reactor volume. Because it's defined by the ideal gas law, changes in the effective temperature caused by turning on the filament array in an isolated reactor can be measured using chamber pressure according to Gay-Lussac's law. In practice, the reactor is not perfectly isolated due to the effect of leak, so we instead use a leak-corrected expression for the effective temperature given by Eq. (17),

$$\frac{T_{eff}}{T_{eff}^0} = \frac{p}{p^0 + \int \frac{dp}{dt}_{leak} dt} \quad (17)$$

where $T_{eff}^0$ is the effective temperature of the reactor during leak rate measurement, $p^0$ is the initial pressure in the reactor with the filament off, $\frac{dp}{dt}_{leak}$ is the rate of pressure change caused by leak into the reactor and $t$ is the time since the filament was turned on. A detailed derivation of Eq. (17) may be found in the supplementary material.

Equation (17) provides a method of determining changes in $T_{eff}$ from pressure changes in the reactor when the power to the filament array is increased in a step-wise manner. Because the leak rate can change slightly when the temperature of the reactor changes, $\frac{dp}{dt}_{leak}$ must be re-measured throughout the experiment. Immediately following a step change in the power to the array when the reactor is equilibrating to a new condition, $\frac{dp}{dt}_{leak}$ cannot be measured, but must instead be updated once the reactor has reached a new equilibrium.

To extract a reliable integrated leak rate from chamber pressure and filament temperature data over time, we iterated through the data systematically. We consider the data in 30 s pieces (6 data points) by iterating through every 30 s period – so if the first period was data points 1-6, the next would be datapoints 2-7. If the difference between the maximum and minimum filament temperature over the 30 s period is less than 2.7 K, we assume the reactor is equilibrated, and update $\frac{dp}{dt}_{leak}$ as the slope of pressure change over the 30 s period determined from a linear regression. The "window size" of 30 s and temperature cutoff of 2.7 K were both chosen to maximize the number of updates to the leak rate, without analyzing too short a time-span, or introducing bias from thermal effects. The leak rates determined using this method are next interpolated over all the conditions which contained greater than 2.7 K spread in temperature and were numerically integrated using a Reimann sum to give the total pressure change due to leak since the beginning of the experiment. This integrated leak rate is used in Eq. (17) to determine the effective temperature.

To model the effective temperature obtained from Eq. (17), we focus on the two-chamber design as that is the most common reactor configuration and matches our system. It is useful to consider the bottom and top chambers separately as volumes $V_1$ and $V_2$, respectively, with the former effectively isolated from the filament and the latter heavily affected by the state of the filament array. Alternate models would be straightforward to derive for other reactor designs. Equation (16) can be re-written in terms of these two volumes.

$$T_{eff} = \frac{V_1 + V_2}{\int \frac{1}{T} dV_1 + \int \frac{1}{T} dV_2} \quad (18)$$

In $V_1$, the temperature is assumed to be constant, so no further analysis is required. While Eq. (9) provides a closed-form approximation for the temperature field surrounding the filaments, it is unnecessarily detailed for determining effective temperature. To obtain a tractable final expression, the filament array is treated as a planar heat source midway between the stage and lid of the top chamber, resulting in a linear temperature profile (neglecting edge effects). By linearizing the equations for heat transfer (Eq. (13)), we obtain Eq. (19),

$$T_{eff} = \cfrac{1}{\cfrac{1-f}{T_{ref}} + \cfrac{f \ln\left(1 + h_{eff}\left(\frac{T_{fil}}{T_{av}} - 1\right)\right)}{h_{eff}(T_{fil} - T_{av})}} \quad (19)$$

where $f = \frac{V_2}{V_1+V_2}$ is the fraction of the total reactor volume that can be accurately modeled using Eq. (9), $T_{ref}$ is the reference temperature, which is assumed to be 40°C based on setpoints for reactor heating, and $h_{eff}$ is defined by Eq. (20).

$$h_{eff} = \cfrac{\left(\frac{H_1 H_2}{H_1 + H_2}\right)\frac{2\pi}{L}}{\ln\left(\frac{L}{2R_{fil}\pi}\right) + \left(\frac{H_1 H_2}{H_1 + H_2}\right)\frac{2\pi}{L} + \frac{b\lambda}{R_{fil}}} \quad (20)$$

Details of the derivation of Eqs. (19) and (20) can be found in the supplementary material.

    To fit the effective temperature to Eq. (19), the fraction $f$ was calculated from the area of the filament array (371 cm²) multiplied by the height of the chamber divided by the total reactor volume. This gave a value of 0.32 for $f$. The distance between filament wires, chamber height, and radius of the filament wire correspond to those used in Section III.D. Eq. (19) was fit to experimental data (from Eq. (17)) by minimizing the mean squared error. The fitting coefficient $\alpha_0$ is reported in Table 4. The accommodation coefficient $\alpha_0$ from Table 4 corresponds excellently with the same value found from the heat transfer analysis in Table 3 for argon. However, the value obtained for IPA is markedly lower. This does not represent a major flaw, or source of error in the model, since the effective temperature is not strongly impacted by the value of the accommodation coefficient.

    Curve fits in argon (Fig. 7 (a-d)) and IPA (Fig. 8 (a-d)) (gases which do not decompose at high temperature), show that Eq. (19) is an acceptable representation of the effective temperature of the vapor phase. Importantly, over a large range of filament

temperature values, the effective temperature changes by only ~5-10%. The notably higher noise in low-pressure data (Fig. 7 (a) and Fig. 8 (a)) is caused by the increased effect of leak at low pressure. Furthermore, the effect of pressure change is minimal. There is some notable deviation from the shape of the model and the shape of the data at temperatures above ~550 K (Fig. 7 and 8) for both IPA and argon. Because these effects are observed in argon gas, which does not absorb infrared radiation, it is likely caused by increased leak or radiative heating of the reactor body which becomes much more pronounced at high filament temperatures due to the quartic scaling of radiative power with filament temperature (Eq. (3)). Accounting for this is outside the scope of work as such high filament temperatures are rarely used.

Table 4. Accommodation coefficient for argon and IPA fit from effective temperature

|  | $\alpha_0$ (non-dim) |
|---|---|
| IPA | 0.064 |
| Argon | 0.16 |

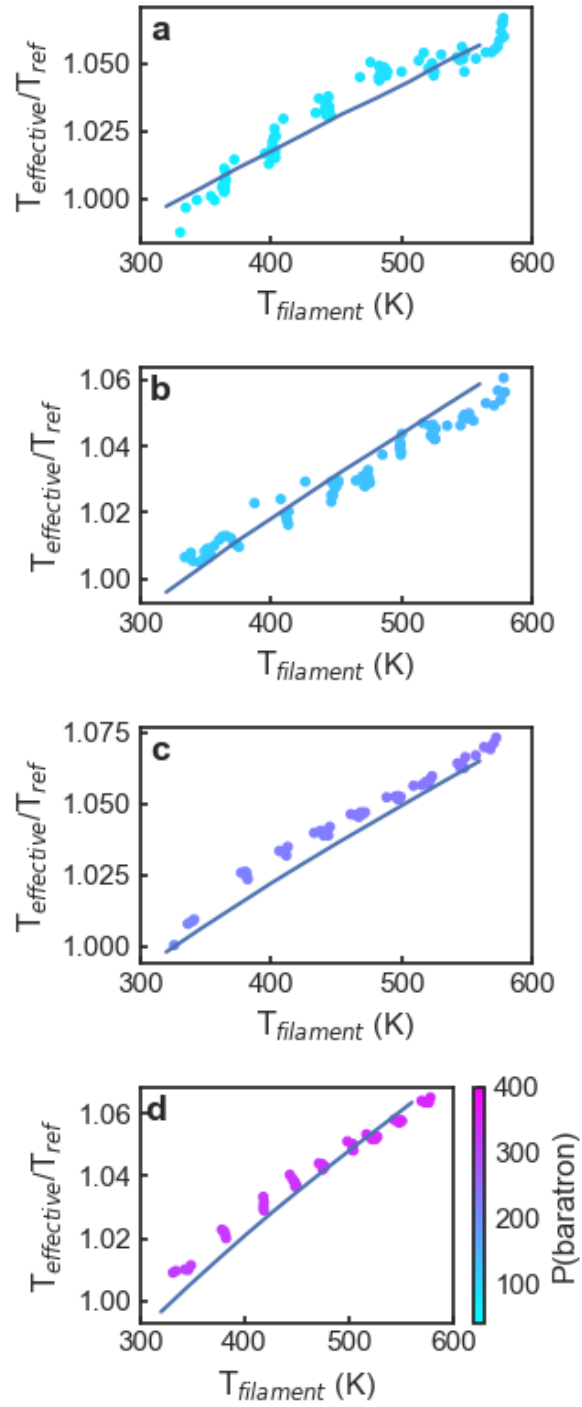

FIG. 7. Ratio of effective temperature to reference temperature versus filament temperature for argon at nominal pressures of (a) 50 mTorr (b) 100 mTorr (c) 200 mTorr and (d) 300 mTorr

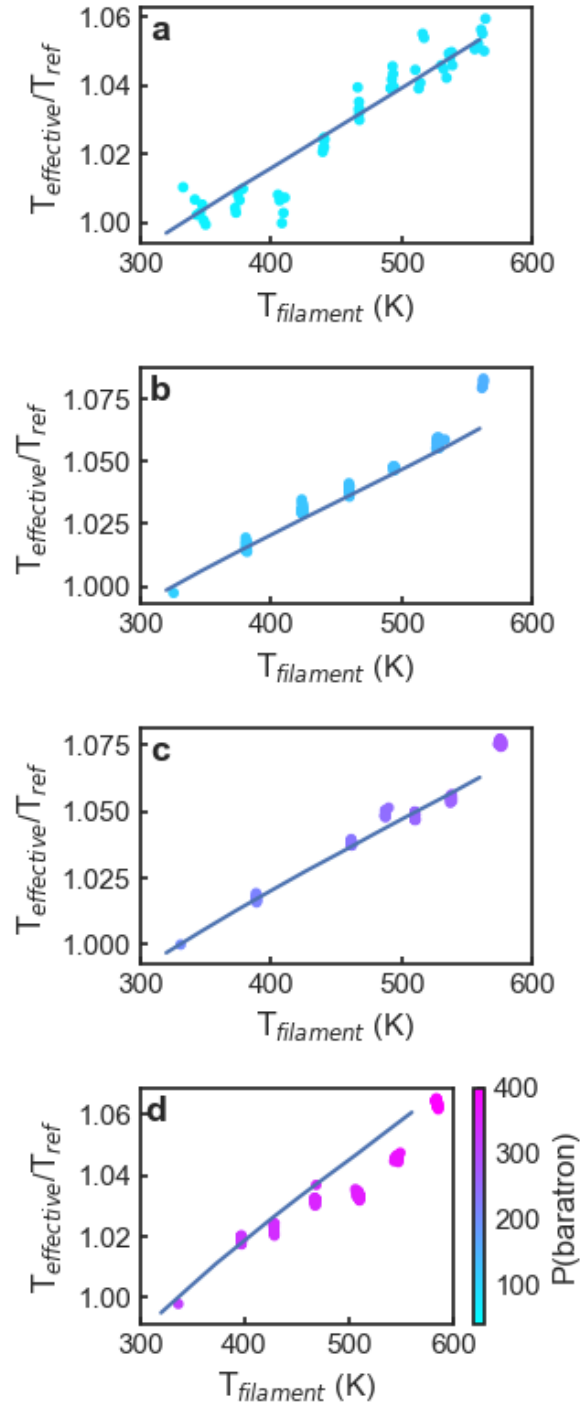

FIG. 8. Ratio of effective temperature to reference temperature versus filament temperature for IPA at nominal pressures of (a) 50 mTorr (b) 100 mTorr (c) 200 mTorr and (d) 300 mTorr

## IV. SUMMARY AND CONCLUSIONS

The published iCVD literature contains very little discussion of the fundamental processes underlying vapor-phase mixing and heat transfer in the reactor. This study attempts to clarify the fundamental processes which underly contacting patterns in the vapor phase, heat transfer from the filament and heat transfer from the reactor body. We present an overview of the dimensionless numbers important to these three cases. Detailed analysis of $Pe_m$ in mass transfer and $Kn$ in heat transfer from the array have been considered. The understanding we develop enable more precise measurements from the iCVD reactor, a concrete understanding of the conditions under which film uniformity or kinetics may be affected by poor mixing, and the temperature profile and heat transfer from the filament array to guide a better understanding of the initiation process, which has largely gone unexplained[13].

For mass transfer, we find that the iCVD reactor is ideally well mixed, and that deviations from perfect mixing can be fully quantified by $Pe_m$. We validate the perspective that under ideal conditions, the reactor is well mixed by performing step-response experiments with IPA and argon using a Pirani gauge for real-time composition sensing, finding that the resulting step response curve matches the design equation for a well-mixed reactor.[21] To help visualize the role of mixing in the reactor, we perform simulations in Autodesk CFD at a range of $Pe_m$ to replicate the step response experiments, finding that measuring gas composition at the outlet of the reactor has poor sensitivity in determining the degree of mixing, but that by placing two sensors at disparate locations in the reactor, comparison of their readout will give strong sensitivity to mixing non-ideality.

We also performed simulations of binary gas mixing at the same range of $Pe_m$ with both a 1-chamber and 2-chamber reactor design. These simulations demonstrated that the mixing was not impacted by the additional chamber, illustrating the utility of simulation in design innovation. Furthermore, these mixing simulations represent the potential effect of poor mixing on an iCVD deposition and provide a clear visualization of what the effect will be at the full range of transitional $Pe_m$. We find that below $Pe_m$ of 0.3 the reactor can be considered as well-mixed based on our simulations. This suggests that film non-uniformities may arise above this point. Additionally, our work illustrates the potential of the Pirani gauge in the characterization of vapor composition, which is simple, low cost, and precise.

For heat transfer, analysis of dimensionless numbers shows that ideal heat transfer to the process gas occurs at low Kn via conduction with no impact from the Knudsen jump. Interfacial heat transfer resistances developing at low pressures can be quantified by Kn. To validate this perspective of ideal heat transfer, we build on past work from Bakker *et al.*[19] to measure heat transfer conducted from the filament wires and find strong agreement with an approximation of the Knudsen jump based on the planar geometry[20], though the system geometry likely causes some deviation for low-pressure argon. Temperature-independent thermal conductivities of 0.00119 and 0.00117 $Wm^{-1}K^{-3/2}$ were obtained for IPA and argon, respectively, compared with values of 0.0011 and 0.00109 $Wm^{-1}K^{-3/2}$ from the literature[42]. Both measured values correspond well with the literature and differ from the literature by a constant factor, indicating that there may be some impact from edge effects or inaccurate dimensions used in our model. While it is important to acknowledge that the model we present makes significant approximations, its ability to describe our data and its correspondence with literature values for gas thermal conductivity suggests that it accurately captures the primary physical drivers of heat transfer to the process vapor in iCVD. For Kn greater than 0.4, non-idealities related to the filament temperature are expected to appear.

While it may appear from our data that heat transfer is strongly affected by pressure at normal iCVD operating pressures, it should be noted that IPA and argon have much smaller collision diameters than normal iCVD process gases, therefore iCVD largely avoids interfacial heat transfer resistances from the Knudsen jump under normal conditions.[19] The heat transfer model presented provides a fundamental understanding of the heat transfer process from the filament arrays, enabling future insight into methods of improving the initiation efficiency during iCVD. The correspondence between predicted thermal conductivity from our model with the literature further supports the treatment of neglecting absorption of thermal radiation by the process vapor, and thermal convection.

To make the application of the ideal gas law rigorous in the iCVD reactor system, we introduce the concept of an effective reactor temperature and provide a model for it. By turning on the filament array while it is closed off from the pump, we measured the change in the effective temperature caused by the filament array, finding that it is predicted well by our model and that its overall effect is limited to a 5-10% change in the reactor pressure. Notably, fitting the effective temperature yielded fitting coefficients (specifically, the accommodation coefficient) of 0.06 and 0.16 for IPA and argon respectively. While the measurement of effective temperature relies on the accommodation coefficient, it is not strongly affected by it, leading to high error in the value obtained. While the accommodation coefficient values obtained from the measurement of diffusive heat transfer in argon matched (0.15) matched well, that for IPA (0.33) was significantly higher. While this difference could be caused by absorption of thermal radiation, further experiments are needed to accurately test such a hypothesis.

Noting the surprising diversity of innovation from efforts to push the feasible boundaries of deposition at high monomer saturation[26–29], we hope that our work inspires and enables the exploration of other boundaries at high $Pe_m$ or Kn. We have left experimental investigation of non-ideality in mass flow at high Kn (termed Knudsen flow), the effects of high $Pe_t$ and a detailed investigation of potential absorption of thermal radiation by the process gas to future work. We also note that further investigations are needed to develop a fundamental understanding of the apparent increase in effective temperature observed at high filament temperature. Additionally, innovative reactor design is needed to obtain uniform films from iCVD at high $Pe_m$. Having introduced the Pirani gauge as a thermal conductivity-based composition sensing device, it is also interesting to consider drawing correlations between the Pirani gauge reading and the Knudsen jump phenomenon at the filament. Finally, it is important to underscore the importance of utilizing the fundamental understanding gained from these experiments in further developing an understanding of the chemical reactions which occur in iCVD by leveraging the temperature profile and well-mixed design equation for computational modeling of reactions in the vapor phase.

# S. SUPPLEMENTARY MATERIAL

## S1. Names of animation files for step response simulations:

https://cornell.box.com/s/i3m84ymi0av0m49u1gf1capgd4cdm96b

4_cm2pers.avi:

Video animation from a simulated step response showing the change in composition over time throughout the reactor using a diffusivity of 4 cm$^2$ s$^{-1}$

16_cm2pers.avi:

Video animation from a simulated step response showing the change in composition over time throughout the reactor using a diffusivity of 16 cm$^2$ s$^{-1}$

64_cm2pers.avi:

Video animation from a simulated step response showing the change in composition over time throughout the reactor using a diffusivity of 64 cm$^2$ s$^{-1}$

678.528_cm2pers.avi:

Video animation from a simulated step response showing the change in composition over time throughout the reactor using a diffusivity of 678.528 cm$^2$ s$^{-1}$

## S2. Details of residence time distribution models
### S2.1 Additional data for Pe = 0.03 simulation

This section contains simulated data from the low Peclet number, high-diffusivity ($D$ = 678.528 cm$^2$ s$^{-1}$) simulation. This data was not included in the main text because it was in-distinguishable from the ideal case. Figure S2.1 shows a plot of this data in black, against the ideal case as red dotted line, showing no discernable difference. The r$^2$ value for the fit is 0.9999986.

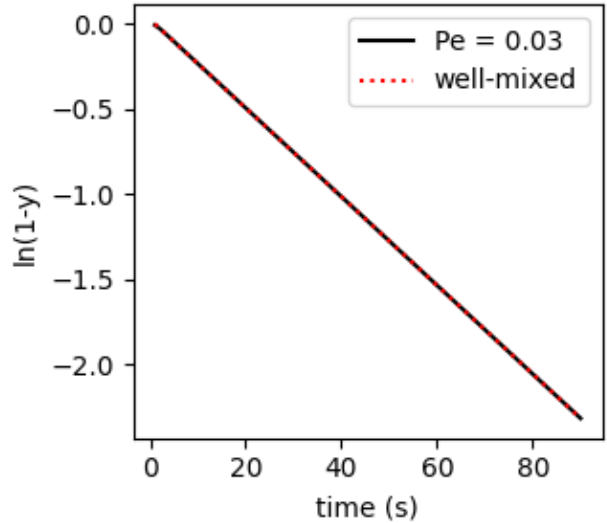

**Figure S2.1.** Comparison between well-mixed assumption and Pe = 0.03 case.

## S2.2 Calculation of reactor volume from residence time distribution

The plot shown in Figure 4 (c) of the main text shows an exponential fit according to the residence time distribution in a well-mixed system[21] given by Eq. (S2.1),

$$RTD \sim \exp\left(-\frac{tF_{in}}{V_r}\right)$$

(S2.1)

Where $F_{in}$ is the volumetric flowrate into the reactor (1.04 sccm), $V_r$ is the volume of the reactor. The ratio of $V_r/F_{in}$ can be determined from a linear regression of ln(RTD), which we show in Fig. S2.2. Note that because of the pressure driven flows at the beginning of the experiment, the y-intercept can be used to determine the total volume of flow accounted for by the pressure driven flows, but that the slope carries physical meaning for the system as a whole. When we perform the regression, we find that the ratio $\frac{F_{in}}{V_r}$ equals 0.0133 s$^{-1}$. For a flowrate of 1.04 sccm, this suggests a reactor volume of 5.7 L, which we report in the main text. First, we convert the molar flow of 1.04 sccm into a volumetric flowrate,

$$F_{in} = 1.04\ sccm * \left(\frac{p^0 T}{pT^0}\right) = 1.04\ sccm * \left(\frac{760\ torr}{0.2\ torr}\frac{315.15\ K}{273.15\ K}\right) = 4560\ cm^3\ min^{-1}$$

$$V_r = \frac{F_{in} min}{0.0133\ s^{-1} 60s} = 5700 cm^3 = 5.7L$$

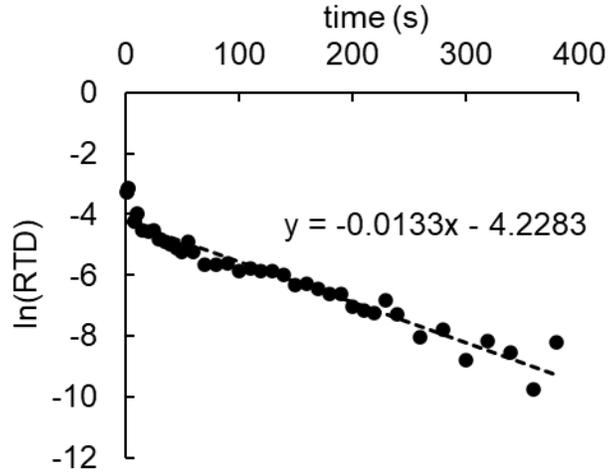

**Figure S2.2.** Linearization of the RTD measurement, showing a fit to data

## S3. Derivation of Eqs. (11) and (12) from the main text

From Su et al.[20], an approximate description for the interfacial heat transfer is given by Eq. (S3.1),

$$T_R = T_{fil} + \frac{\zeta_T \mu}{p}\sqrt{\frac{2k_B T_{fil}}{m}}\,\mathbf{n} \cdot \nabla T$$

(S3.1)

where $T_R$ corresponds to $T_e$ in Su *et al.* and is the extrapolated temperature of the gas molecules at the surface of the filament array, $T_{fil}$ is the temperature of the filament itself, and corresponds to $T_w$ in Su *et al.* (see figure S3.1), $p$ is the pressure, $k_B$ is Boltzmann's constant, $m$ is the mass of a gas molecule, $\mu$ is the dynamic viscosity of the gas, $\mathbf{n}$ is the unit normal vector to the filament surface, and $\zeta_T$ is the temperature jump coefficient defined by Eq. (S3.2).

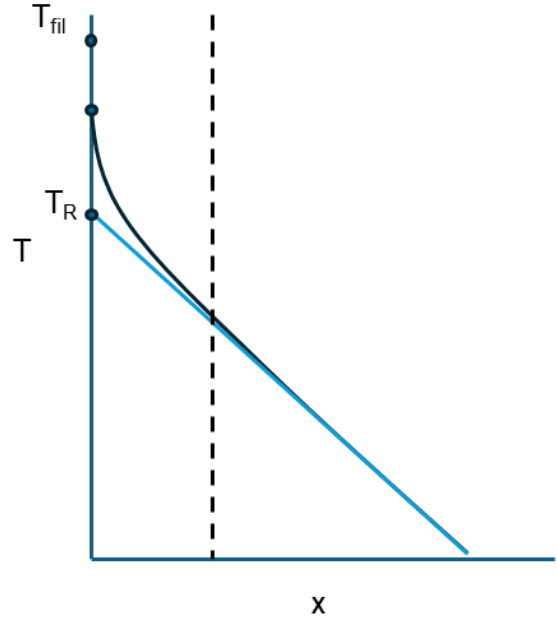

**Figure S3.1.** Temperature profile in the Knudsen layer, showing temperature profile close to the array including $T_R$, the extrapolated temperature at the surface of the array, $T_{fil}$, the temperature of the filament and a dotted line showing the approximate width of the Knudsen layer.

$$\zeta_T = \frac{\gamma\sqrt{\pi}}{(\gamma+1)Pr}\frac{2-\alpha_0}{\alpha_0}$$

(S3.2)

where $\gamma$ is the heat capacity ratio, having a value of 1.105 for IPA[51] and 1.66 for argon, $Pr$ is the Prandtl number, defined as $Pr = \frac{c_p\mu}{m\kappa}$, where $c_p$ is the heat capacity of a molecule, $\mu$ is the dynamic viscosity, $\kappa$ is the thermal conductivity, and $m$ is the mass of a molecule, and $\alpha_0$ is the accommodation coefficient, representing the fraction of molecules which are diffusely reflected at the wall. We note that the temperature gradient at the surface of the filament can be written in terms of the heat transfer from the filament using Eq. (S3.3),

$$\frac{\vec{j}_q}{\kappa} = -\nabla T$$

(S3.3)

where $j_q$ is heat flux (W m$^{-2}$). This can further be converted to Eq. (S3.4) by dotting with the surface unit normal and noting that $j_q = \frac{q_{fil}}{2\pi R_{fil}} n$, where $q_{fil}$ is the total power conducted away from the filament per unit length in W m$^{-1}$, $R_{fil}$ is the radius of the filament. This is amenable to substitution into Eq. (S3.1).

$$\frac{q_{fil}}{2\pi R_{fil} \kappa} = -n \cdot \nabla T$$

(S3.4)

Substitution of Eqs. (S3.2)-(S3.4) into Eq. (S3.1) yields Eq. (S3.5),

$$T_R = T_{fil} - \frac{\gamma m \sqrt{\pi}}{p(\gamma+1)c_p} \frac{2-\alpha_0}{\alpha_0} \sqrt{\frac{2k_B T_{fil}}{m}} \frac{q_{fil}}{2\pi R_{fil}}$$

(S3.5)

We now multiply the equation by $T_{fil}^{1/2}$, to get Eq. (S3.6).

$$T_R T_{fil}^{\frac{1}{2}} = T_{fil}^{\frac{3}{2}} - \frac{\gamma m \sqrt{\pi}}{(\gamma+1)c_p} \frac{2-\alpha_0}{\alpha_0} \frac{T_{fil}}{p} \sqrt{\frac{2k_B}{m}} \frac{q_{fil}}{2\pi R_{fil}}$$

(S3.6)

For temperature jump of less than 50°C in filament temperatures over 200°C, error is less than ~5% when we assume $T_R T_{fil}^{\frac{1}{2}} \sim T_R^{\frac{3}{2}}$ to get Eq. (S3.7).

$$T_R^{\frac{3}{2}} = T_{fil}^{\frac{3}{2}} - \frac{\gamma m \sqrt{\pi}}{(\gamma+1)c_p} \frac{2-\alpha_0}{\alpha_0} \frac{T_{fil}}{p} \sqrt{\frac{2k_B}{m}} \frac{q_{fil}}{2\pi R_{fil}}$$

(S3.7)

To make Eq. (S3.7) consistent in usage of the heat transfer potential, we substitute Eq. (S3.8) for the temperature,

$$\frac{\phi}{-\frac{2}{3}\kappa'} = T^{\frac{3}{2}}$$

(S3.8)

where $\phi$ is the heat transfer potential, $\kappa'$ is the temperature independent thermal conductivity defined as $\kappa' = \frac{\kappa}{\sqrt{T}}$. Where thermal conductivity of an ideal gas from kinetic theory of gases is given by Eq. (S3.9),

$$\kappa = \frac{f\sqrt{T}}{3d^2}\sqrt{\frac{k_B^3}{\pi^3 m}}$$

(S3.9)

where $f$ is the degrees of freedom of a molecule and $d$ is the collision diameter of the molecule. Substitution of Eqs. (S3.8) and (S3.9) yield Eq. (S3.10).

$$\phi_{fil} - \phi_R = -\frac{fk_B^2\sqrt{8}}{9d^2\pi}\frac{\gamma}{(\gamma+1)c_p}\left(\frac{2-\alpha_0}{\alpha_0}\right)\frac{T_{fil}}{p}\frac{q_{fil}}{2\pi R_{fil}}$$

(S3.10)

Next, using the definition of the mean free path, we can eliminate the explicit dependence on temperature and pressure in Eq. (S3.10) using Eq. (S3.11), to get Eq. (S3.12).

$$\frac{\lambda\sqrt{2}\pi d^2}{k_B} = \frac{T}{p}$$

(S3.11)

$$\phi_{fil} - \phi_R = -\frac{4fk_B}{9}\frac{\gamma}{(\gamma+1)c_p}\left(\frac{2-\alpha_0}{\alpha_0}\right)\lambda\frac{q_{fil}}{2\pi R_{fil}}$$

(S3.12)

Finally, we can simplify the heat capacity as $c_p = \gamma c_v = \gamma\frac{fk_B}{2}$, which gives Eq. (S3.13).

$$\phi_{fil} - \phi_R = -\frac{8}{9}\frac{1}{(\gamma+1)}\left(\frac{2-\alpha_0}{\alpha_0}\right)\lambda\frac{q_{fil}}{2\pi R_{fil}}$$

(S3.13)

By comparison with Eq. (11) from the main text (Eq. S3.14), we obtain the parameter $b$ given by Eq. (12) in the main text, and Eq. (S3.15) here.

$$\phi_{fil} - \phi_R = -\frac{b\lambda q_{fil}}{2\pi R_{fil}}$$

(S3.14)

$$b = \frac{8}{9}\frac{2-\alpha_0}{\alpha_0(\gamma+1)}$$

(S3.15)

## S4. Derivation of Model for Heat Transfer from the Filament

Conduction heat transfer in a gas follows Eq. (S4.3). At steady-state, conservation of energy requires:

$$\nabla \kappa \nabla T = 0$$

Because $\kappa \sim \sqrt{T}$ from the kinetic theory of gases,[33] the equation governing heat transfer can be written as Laplace's equation,

$$\nabla^2 \phi = 0$$

(S4.1)

where we introduce Eq. (7) from the main text as Eq. (S4.2):

$$\phi = -\frac{2}{3}\kappa' T^{\frac{3}{2}}$$

(S4.2)

$$\kappa' = \frac{\kappa}{\sqrt{T}} = \frac{f}{3d^2}\sqrt{\frac{k_b^2}{\pi^3 m}}$$

(S4.3)

To solve Eq. (S4.1) in the iCVD geometry it is helpful to consider that spacing between the filaments is small compared to the distance to the stage, and they can therefore be approximated as an infinite line of equally spaced 2D point sources (line sources) of equal magnitude. This is if we ignore edge effects in the reactor.

The Cauchy-Reimann equations show that the real (and imaginary) part of any complex function is a solution of Laplace's equation.

Using this fact, any 2D solution of Laplace's equation expressed in terms of circular harmonics, or Green's functions (point sources, sinks, dipoles, quadrupoles etc.), consists of finding a complex function with the right types of singularities in the right places. If a complex function is found with the appropriate singularities, then it is the solution.

To obtain a solution to our problem, it is first necessary to consider the complex natural logarithm. Using the polar representation for the complex plane, $z = re^{i\theta}$, it can be shown that the real part of the complex natural log corresponds to the natural log of the absolute value of its argument.

$$Re(\ln(z)) = \ln|z| = \ln(r)$$

(S4.4)

From this, we can see already that $Re(\ln(z))$ corresponds to the solution of Laplace's equation with a point source at the origin. Since we're interested in a line of point sources, we're only interested in logarithmic singularities.

Extending Eq. (S4.4), we can see that the real part of the natural log of a complex function corresponds to the natural log of the absolute value of the function.

$$Re(\ln(f(z))) = \ln|f(z)|$$

The natural log function goes to negative infinity as its argument goes to zero and infinity as its argument goes to infinity, but is continuous everywhere else. Therefore, singular values of $\ln|f(z)|$ are places where $|f(z)|$ are either zero or infinity.

Because $\ln(x)$ goes to negative infinity at zero, and infinity at infinity, zeros of $|f(z)|$ correspond to point sinks, and poles of $|f(z)|$ correspond to point sources, or vice versa for $-\ln(x)$. Places where the limiting value of $|f(z)|$ is zero from one direction, and infinity from another correspond to higher order harmonics. For instance, $|\exp(1/z)|$ corresponds to a dipole at the origin.

Finding the solution of Laplace's equation for a line of equally-spaced point sources of equal magnitude therefore consists of finding a periodic function with a line of equally spaced zeros. The only zeros of $\sin(z)$ are located at $n\pi$, so we expect it matches the solution to our problem.

$$Re(\ln(\sin(z))) = \ln|\sin(z)|$$

Note that in the limit approaching zero, $\ln|\sin(z)|$ behaves as $\ln(r)$, which we know to be the solution for a point source.

$$\lim_{z \to 0} \ln|\sin(z)| = \ln(|z|) = \ln(r)$$

(S4.5)

Because ln(sin(z)) is periodic, this means that indeed $\ln|\sin(z)|$ corresponds to an infinite line of equally spaced point sources of equal magnitude. Other trigonometric functions yield similar solutions, either with alternating point sources and point sinks, point sinks instead of point sources, or with the sources shifted on the *x*-axis.

It can also be shown that ln|sin(z)| has appropriate behavior for large *y*, by considering the following limit:

$$\ln|\sin(x+iy)| = \ln\left|\frac{e^{i(x+iy)} - e^{-i(x+iy)}}{2i}\right|$$

$$\lim_{y\to\infty} \ln|\sin(z)| = \ln\left|-\frac{e^y}{2i}\right|$$

$$\lim_{y\to\infty} \ln|\sin(z)| = \ln\left|i\frac{e^y}{2}\right| = y - \ln(2)$$

(S4.6)

This is the correct behavior for large values of *y*, so we consider that the solution of Laplace's equation for an infinite line of equally spaced point sources of equal magnitude is given by Eq. (S4.7).

$$\phi \sim Re(\ln(\sin(z)))$$

(S4.7)

To make this a solution to a temperature field for the iCVD reactor, we introduce a proportionality constant ($c_2$) which will be related to the total flow of heat from the filament, and re-scale *z* so that the periodicity of sin(*z*) matches the spacing of the filament (*L*).

$$\phi = c_1 + c_2 Re\left(\ln\left(\sin\left(\frac{z\pi}{L}\right)\right)\right)$$

To convert complex numbers to *x,y*-coordinates, we substitute $z = x + iy$.

$$\phi = c_1 + c_2 Re\left(\ln\left(\sin\left(\frac{x\pi}{L} + \frac{iy\pi}{L}\right)\right)\right)$$

Next, we re-write sin(*z*) as a complex function.

$$\phi = c_1 + c_2 Re\left(\ln\left(\sin\left(\frac{x\pi}{L}\right)\cosh\left(\frac{y\pi}{L}\right) + i\cos\left(\frac{x\pi}{L}\right)\sinh\left(\frac{y\pi}{L}\right)\right)\right)$$

As discussed above, the real part of ln(*z*) is equal to ln(|*z*|), which allows for simplification. We will also appeal to the principle of linear superposition to add the

homogenous temperature profile resulting from a heat flux of $q_{stage}$ from the stage to the lid when the filament is off (which is linear in $y$).

$$\phi = c_1 + c_2 \ln\left(\sqrt{\sin^2\left(\frac{x\pi}{L}\right)\cosh^2\left(\frac{y\pi}{L}\right) + \cos^2\left(\frac{x\pi}{L}\right)\sinh^2\left(\frac{y\pi}{L}\right)}\right) + c_3 y$$

(S4.8)

Next, we need to solve for $c_1$, $c_2$ and $c_3$ using boundary conditions. Normally, we only specify 2 boundary conditions, however, because we have two limiting cases given by Eqs. (S4.5) and (S4.6) which have little impact on each other, we specify 2 boundary conditions for each limiting condition.

First, we consider the region close to the surface of the filament array given by Eq. (S4.5), which approximates Eq. (S4.8) as Eq. (S4.9). Note that because we scaled 'z' by $\frac{L}{\pi}$ to match the filament spacing, the argument inside the natural log is scaled.

$$\phi = c_1 + c_2 \ln\left(\frac{r\pi}{L}\right) + c_3 y$$

(S4.9)

First, we have the energy-balance integral condition at the surface of the filament,

$$\int_0^{2\pi} j_{u\,r=R_{fil}} R_{fil} d\theta = q_{fil}$$

(S4.10)

We plug in Eq. (S4.9) and solve to obtain a value for $c_2$.

$$\left(\frac{\partial \phi}{\partial r}\right)_{r=R_{fil}} = j_{q\,r=R_{fil}} = \frac{c_2}{R_{fil}} + c_3 \sin(\theta)$$

$$\int_0^{2\pi} c_2 + c_3 R_{fil} \sin(\theta)\, d\theta = q_{fil}$$

$$c_2 = \frac{q_{fil}}{2\pi}$$

Next, we note that the filament radius is small enough that the temperature at the surface of the filament is approximately constant, giving the second boundary condition as Eq. (S4.11).

$$\phi(R_{fil}) = \phi_R$$

(S4.11)

We solved this by plugging in Eq. (S4.9) to find $c_1$.

$$c_1 = \phi_R - \frac{q_{fil}}{2\pi} \ln\left(\frac{R_{fil}\pi}{L}\right)$$

We can plug these into Eq. (S4.8) to obtain Eq. (S4.12)

$$\phi = \phi_R - \frac{q_{fil}}{2\pi} \ln\left(\frac{R_{fil}\pi}{L}\right) + \frac{q_{fil}}{2\pi} \ln\left(\sqrt{\sin^2\left(\frac{x\pi}{L}\right)\cosh^2\left(\frac{y\pi}{L}\right) + \cos^2\left(\frac{x\pi}{L}\right)\sinh^2\left(\frac{y\pi}{L}\right)}\right) + c_3 y$$

(S4.12)

Just as we considered the region close to the filament to find $c_1$ and $c_2$, we will consider the region far from the filament (where the temperature profile is linear) to find $c_3$. Now we'll consider the solution in the limit of Eq. (S4.12) as |y| goes to infinity (given by Eq. (S4.6)) to give Eq. (S4.13).

$$\lim_{|y|\to\infty} \phi = \phi_R - \frac{q_{fil}}{2\pi} \ln\left(\frac{R_{fil}\pi}{L}\right) + \frac{q_{fil}}{2\pi}\left(\left|\frac{y\pi}{L}\right| - \ln(2)\right) + c_3 y$$

(S4.13)

Eq. (S4.13) is a good approximation when $y$ is on the order of $H_1$ and $H_2$. We now apply the boundary condition at the stage $\phi(-H_2) = \phi_{stage}$ to obtain Eq. (S4.14),

$$\phi_{stage} = \phi_R - \frac{q_{fil}}{2\pi} \ln\left(\frac{R_{fil}\pi}{L}\right) + \frac{q_{fil}}{2\pi}\left(\frac{H_2\pi}{L} - \ln(2)\right) - c_3 H_2$$

(S4.14)

and $\phi(H_1) = \phi_{lid}$ to obtain Eq. (S4.15).

$$\phi_{lid} = \phi_R - \frac{q_{fil}}{2\pi} \ln\left(\frac{R_{fil}\pi}{L}\right) + \frac{q_{fil}}{2\pi}\left(\frac{H_1\pi}{L} - \ln(2)\right) + c_3 H_1$$

(S4.15)

We subtract Eqs. (S4.14) from (S4.15) to solve for $c_3$,

$$\phi_{lid} - \phi_{stage} = \frac{q_{fil}}{2L}(H_1 - H_2) + c_3(H_1 + H_2)$$

$$\frac{\phi_{lid} - \phi_{stage}}{H_1 + H_2} - \frac{q_{fil}}{2L}\frac{H_1 - H_2}{H_1 + H_2} = c_3$$

Now, we can plug in $c_3$ to Eq. (S4.12) to give Eq. (S4.16), which describes the overall temperature profile.

$$\phi = \phi_R + \frac{q_{fil}}{2\pi} \ln\left(\frac{L}{R_{fil}\pi}\sqrt{\sin^2\left(\frac{x\pi}{L}\right)\cosh^2\left(\frac{y\pi}{L}\right) + \cos^2\left(\frac{x\pi}{L}\right)\sinh^2\left(\frac{y\pi}{L}\right)}\right)$$
$$+ \left(\frac{\phi_{lid} - \phi_{stage}}{H_1 + H_2} - \frac{q_{fil}}{2L}\frac{H_1 - H_2}{H_1 + H_2}\right)y$$

(S4.16)

Next, we want to use this temperature field to obtain the overall thermal conductivity of the vapor phase. First, we assume that the lid and stage temperatures are equal. Since we're interested in the values of $\phi$ at the lid and the stage, we use the limit from Eq. (S4.6) to evaluate Eq. (S4.16) at $y = -H_2$, giving Eq. (S4.17).

$$\frac{(\phi_{stage} - \phi_R)}{q_{fil}} = \frac{1}{2\pi}\ln\left(\frac{L}{2R_{fil}\pi}\right) + \left(\frac{H_1 H_2}{H_1 + H_2}\right)\frac{1}{L}$$

(S4.17)

Eq. (S4.7) is provided in the main text as Eq. (10). Note that Eq. (S4.17) is combined with Eq. (S3.14) to give Eq. (S4.18), which is Eq. (13) from the main text.

$$\phi_{fil} = \phi_{stage} - \frac{q_{fil}}{2\pi}\left(\ln\left(\frac{L}{2R_{fil}\pi}\right) + \left(\frac{H_1 H_2}{H_1 + H_2}\right)\frac{2\pi}{L} + \frac{b\lambda}{R_{fil}}\right)$$

(S4.18)

## S5. Derivation of Model for the effective temperature
### S5.1 Reactor Effective Temperature definition

Assume that the gas locally follows the ideal gas law, and that temperature varies spatially in the reactor:

$$T = f(x) = \frac{pv}{R_G}$$

Based on conservation of momentum, we consider that there are no pressure gradients in the reactor, or that those associated with flow are negligible:

$$p = RT\frac{dn}{dV} = constant$$

$$\int \frac{dn}{dV}dV = n$$

$$\int \frac{p}{RT} dV = n$$

$$\int \frac{1}{T} dV = \frac{nR_G}{p}$$

This tells us that there is an average temperature which is defined by one over the average of $1/T$ over the full volume of the reactor. This is the temperature of the reactor for which the ideal gas law will hold.

$$T_{eff} = \frac{V}{\int \frac{1}{T} dV}$$

(S5.1)

## S5.2 Measuring Effective Temperature (batch)

For the simple case of a leaky reactor with no reaction, we can write:

$$T_{eff} = \frac{pV}{(n_0 + \dot{n}_{leak} t) R_G}$$

(S5.2)

At constant temperature:

$$n_0 + \dot{n}_{leak} t = p \frac{V}{R_G T_{eff,leak}}$$

Take the derivative with respect to time to isolate the leak-rate:

$$\dot{n}_{leak} = \frac{V}{R_G T_{eff,leak}} \frac{dp}{dt}\bigg|_{leak}$$

(S5.3)

Subbing Eq (S5.3) into Eq (S5.2):

$$T_{eff} = \frac{p}{\frac{p_o}{T_{eff,0}} + \frac{dp}{dt}\big|_{leak} \frac{t}{T_{eff,leak}}}$$

$$T_{eff} = \frac{p T_{eff,leak}}{p_o + \frac{dp}{dt}\big|_{leak} t}$$

(S5.4)

## S5.3 Reactor Effective Temperature Model

Take the definition of the effective temperature and break it up into three volumes according to Eq. (S5.5).

$$T_{eff} = \frac{V}{\int \frac{1}{T} dV} = \frac{V_1 + V_2 + V_3}{\int \frac{1}{T} dV_1 + \int \frac{1}{T} dV_2 + \int \frac{1}{T} dV_3}$$

(S5.5)

$V_1$ is the volume of the bottom chamber of the reactor, which is effectively isolated from the filament, and remains at some low reference temperature throughout the experiment. $V_2$ is the volume of the reactor that is affected by filament heating, which is well modeled by Eq. (S4.16) and makes up most of the top chamber where the filament array is located. $V_3$ is the part of the reactor in the top chamber which is not well modeled by Eq. (S4.16), and accounts for the edges of the chamber. This volume has not been mentioned in the main text, as we lump it with $V_1$, however we include it here to show how a more general expression might be obtained.

First, we assume that $V_1$ is at a constant temperature without thermal gradients:

$$\int \frac{1}{T} dV_1 = \frac{V_1}{T_{ref}}$$

(S5.6)

While the temperature in $V_2$ is accurately described by Eq. (S4.16), for the purposes of calculating the effective temperature, it is expedient to treat the filament array as a planar heat source, such that the temperature profile is linear. Again, we assume the filament is located midway between the lid of the reactor and the reactor stage. Furthermore, even though heat transfer in a gaseous medium tends to go as $T^{3/2}$, instead of $T$ over the temperature range of interest (20-200°C), approximating it as $\sim T$ yields a very close approximation.

We will approximate most of the thermal gradients in the reactor with linear temperature profiles, so we will first evaluate Eq. (S5.5) for an arbitrary linear profile in the $y$-direction. First we define the temperature profile,

$$T = a_1 + a_2 y$$

Invert it,

$$\frac{1}{T} = \frac{1}{a_1 + a_2 y}$$

Integrate the inverted temperature profile over volume substituting $u = a_1 + a_2 y$,

$$A \int_{Y_1}^{Y_2} \frac{1}{T} dy = A \frac{1}{a_2} \int_{\frac{a_1}{a_2}+Y_1}^{\frac{a_1}{a_2}+Y_2} \frac{1}{u} du$$

$$A \int_{Y_1}^{Y_2} \frac{1}{T} dy = A \frac{1}{a_2} [\ln(u)]_{\frac{a_1}{a_2}}^{\frac{a_1}{a_2}+Y}$$

where $A$ and $Y_2-Y_1$ are area and length such that the total volume being integrated over is $V = A(Y_2 - Y_1)$,

$$A \int_{Y_1}^{Y_2} \frac{1}{T} dy = \frac{V \ln\left(\frac{T(Y_2)}{T(Y_1)}\right)}{T(Y_2) - T(Y_1)}$$

(S5.7)

To solve for $V_3$, we let Eq. (S5.7) integrate from $-H_1$ to $H_2$, and substitute $T_{stage}$ and $T_{lid}$ for the associated temperatures noting that $(H_1+H_2)A_3 = V_3$, where $A_3$ is area of the footprint taken up by $V_3$.

$$\int_{-H_2}^{H_1} \frac{1}{T} dV_3 = \frac{V_3 \ln\left(\frac{T_{lid}}{T_{stage}}\right)}{T_{lid} - T_{stage}}$$

We can note that, when $T_{stage}$ and $T_{lid}$ are within 20 or 30°C, the result can be further simplified to Eq. (S5.8),

$$\int_{-H_2}^{H_1} \frac{1}{T} dV_3 = \frac{V_3}{T_{av}}$$

(S5.8)

where we have defined an average temperature as $T_{av} = \frac{T_{stage}+T_{lid}}{2}$. Next, we consider the plane source in $V_2$. This is identical to the limit of Eq. (S4.16) for large $y$, described by Eq. (S4.6), which gives Eq. (S5.9).

$$\phi = \phi_R + \frac{q_{fil}}{2\pi} \ln\left(\frac{L}{R_{fil}\pi}\right) + \frac{q_{fil}}{2\pi}\left(\left|\frac{y\pi}{L}\right| - \ln(2)\right) + \left(\frac{\phi_{lid} - \phi_{stage}}{H_1 + H_2} - \frac{q_{fil}}{2L}\frac{H_1 - H_2}{H_1 + H_2}\right)y$$

(S5.9)

We can apply Eq. (S4.14) to convert $\phi R$ to $\phi fil$ giving Eq. (S5.10).

$$\phi = \phi_{fil} + \frac{q_{fil}}{2\pi}\left(\frac{b\lambda}{R_{fil}} + \ln\left(\frac{L}{R_{fil}\pi}\right) + \left|\frac{y\pi}{L}\right| - \ln(2) \right.$$
$$\left. + \left(\frac{2\pi(\phi_{lid} - \phi_{stage})}{q_{fil}(H_1 + H_2)} - \frac{\pi}{L}\frac{H_1 - H_2}{H_1 + H_2}\right)y\right)$$

(S5.10)

We note that this Eq. (S5.10) is linear in 'y', therefore we can integrate it piecewise using Eq. (S5.7). First, we substitute -$H_1$ and 0 for $Y_1$ and $Y_2$ respectively, noting that $T(-H_2) = T_{stage}$. Then add a second profile where we substitute 0 and $H_2$ for $Y_1$ and $Y_2$ respectively, noting that $T(H_1) = T_{lid}$. Adding together these two terms we obtain Eq. (S5.11)

$$\int \frac{1}{T} dV_2 = \frac{A_2 H_2 \ln\left(\frac{T(0)}{T_{stage}}\right)}{T(0) - T_{stage}} + \frac{A_2 H_1 \ln\left(\frac{T(0)}{T_{lid}}\right)}{T(0) - T_{lid}}$$

(S5.11)

where $A_2$ is the area of the footprint taken up by $V_2$ In a similar way as for $V_3$, $V_2$ can be approximated with good accuracy when the stage and lid temperatures are within 20-30°C noting that $A_2(H_1+H_2)$ is equal to $V_2$.

$$\int \frac{1}{T} dV_2 = \frac{V_2 \ln\left(\frac{T(0)}{T_{av}}\right)}{T(0) - T_{av}}$$

(S5.12)

$T(0)$ can be calculated from Eq. (S5.10) as Eq. (S5.13), substituting Eq. (S4.2) to convert heat transfer potential to temperature

$$T(0)^{\frac{3}{2}} = T_{fil}^{\frac{3}{2}} - \frac{3q_{fil}}{\kappa' 4\pi}\left(\frac{b\lambda}{R_{fil}} + \ln\left(\frac{L}{2R_{fil}\pi}\right)\right)$$

(S5.13)

From our study of the heat transfer from the filament, we know that heat transfer depends on pressure according to Eq. (S4.18) which can be converted to temperature using the definition of the heat transfer potential (Eq. (S4.2)) to give Eq. (S5.14).

$$q_{fil} = \frac{4\pi\kappa'\left(T_{fil}^{\frac{3}{2}} - T_{av}^{\frac{3}{2}}\right)}{3\left(\ln\left(\frac{L}{2R_{fil}\pi}\right) + \left(\frac{H_1 H_2}{H_1 + H_2}\right)\frac{2\pi}{L} + \frac{b\lambda}{R_{fil}}\right)}$$

(S5.14)

Plug Eq. (S5.14) into (S5.13) to give Eq. (S5.15)

$$T(0)^{\frac{3}{2}} = T_{av}^{\frac{3}{2}} + \frac{\left(\frac{H_1 H_2}{H_1 + H_2}\right)\frac{2\pi}{L}\left(T_{fil}^{\frac{3}{2}} - T_{av}^{\frac{3}{2}}\right)}{\ln\left(\frac{L}{2R_{fil}\pi}\right) + \left(\frac{H_1 H_2}{H_1 + H_2}\right)\frac{2\pi}{L} + \frac{b\lambda}{R_{fil}}}$$

(S5.15)

Only marginal error is incurred by linearizing as:

$$T(0) = T_{av} + \frac{\left(\frac{H_1 H_2}{H_1 + H_2}\right)\frac{2\pi}{L}(T_{fil} - T_{av})}{\ln\left(\frac{L}{2R_{fil}\pi}\right) + \left(\frac{H_1 H_2}{H_1 + H_2}\right)\frac{2\pi}{L} + \frac{b\lambda}{R_{fil}}}$$

Which we plug into Eq. (S5.12),

$$\int \frac{1}{T} dV_2 = \frac{V_2 \ln\left(1 + h_{eff}\left(\frac{T_{fil}}{T_{av}} - 1\right)\right)}{h_{eff}(T_{fil} - T_{av})}$$

(S5.16)

where we define an effective heat transfer coefficient from Eq. (S5.17)

$$h_{eff} = \frac{\left(\frac{H_1 H_2}{H_1 + H_2}\right)\frac{2\pi}{L}}{\ln\left(\frac{L}{2R_{fil}\pi}\right) + \left(\frac{H_1 H_2}{H_1 + H_2}\right)\frac{2\pi}{L} + \frac{b\lambda}{R_{fil}}}$$

(S5.17)

To obtain an expression for the overall effective temperature in the reactor, we substitute Eq. (S5.16) into Eq. (S5.5).

$$T_{eff} = \frac{V_1 + V_2 + V_3}{\frac{V_1}{T_{ref}} + \frac{V_2 \ln\left(1 + h_{eff}\left(\frac{T_{fil}}{T_{av}} - 1\right)\right)}{h_{eff}(T_{fil} - T_{av})} + \frac{V_3}{T_{av}}}$$

(S5.18)

When the stage, lid and reference temperature are all within 30ºC or so, we can write Eq. (S5.19).

$$T_{eff} = \cfrac{1}{\cfrac{1-f}{T_{ref}} + \cfrac{f \ln\left(1 + h_{eff}\left(\frac{T_{fil}}{T_{av}} - 1\right)\right)}{h_{eff}(T_{fil} - T_{av})}}$$

(S5.19)

We have now derived Eq. (15) from the main text.

## ACKNOWLEDGMENTS

The project was sponsored by the National Science Foundation Faculty Early Career Development Program (CMMI-2144171), as well as the Department of the Navy, Office of Naval Research under ONR award N00014-23-1-2189.

## AUTHOR DECLARATIONS

**Conflicts of Interest**

The authors have no conflicts of interest to disclose.

## DATA AVAILABILITY

The data that supports the findings in this study are available upon reasonable request from the corresponding author.